\documentclass[journal]{IEEEtran}
\usepackage[utf8]{inputenc}
\usepackage[T1]{fontenc}
\usepackage{url}
\usepackage{ifthen}
\usepackage{cite}

\usepackage{graphicx,epstopdf,amssymb,amsthm}
\usepackage{color}
\usepackage{epstopdf}
\usepackage[utf8]{inputenc}
\usepackage[english]{babel}
\usepackage{caption}

\usepackage[dvipsnames]{xcolor}

\newcommand{\phil}[1]{\textcolor{black}{#1}}
\newcommand{\baike}[1]{\textcolor{black}{#1}}
\newcommand{\bshe}[1]{\textcolor{black}{#1}}
\newcommand{\bs}[1]{\textcolor{black}{#1}}
\newcommand{\bsheup}[1]{\textcolor{black}{#1}}
\newcommand{\bsheupp}[1]{\textcolor{black}{#1}}

\newcommand{\R}{{\rm I\!R}}

\newtheorem{theorem}{Theorem}
\newtheorem{lemma}{Lemma}
\newtheorem{proposition}{Proposition}
\newtheorem{corollary}{Corollary}
\newtheorem{assumption}{Assumption}
\newtheorem{remark}{Remark}

\newcommand{\diag}{\text{diag}}

\newtheorem{definition}{Definition}
\usepackage[T1]{fontenc}
\usepackage{color}
\usepackage{amsmath}
\usepackage{amsthm}
\usepackage{amssymb}
\usepackage{stmaryrd}
\usepackage{graphicx}
\usepackage{esint}

\makeatletter

\begin{document}
%
\title{On a Networked SIS Epidemic Model with Cooperative and Antagonistic Opinion Dynamics}


\author{
Baike She, 
Ji Liu, 
Shreyas Sundaram, and 
Philip E. Par\'{e}*
\thanks{*Baike She, Shreyas Sundaram, and Philip E. Par\'{e} are with the Elmore Family School of Electrical and Computer Engineering, Purdue University.
Ji Liu is with the Department of Electrical and Computer Engineering, Stony Brook University. 
E-mails: \{bshe, sundara2, philpare\}@purdue.edu.
Research supported in part by the C3.ai Digital Transformation Institute sponsored by C3.ai Inc. and the Microsoft Corporation, and in part 
   by the National Science Foundation, grants NSF-CMMI \#1635014
   and NSF-ECCS \#2032258.
}
}

%



\maketitle
\begin{abstract}
We
propose a
mathematical model to study coupled epidemic and opinion dynamics in a network of communities. Our model captures SIS epidemic dynamics whose evolution is \phil{dependent on}
the opinions of the communities toward the epidemic, and vice versa. 
In particular, we allow both cooperative and antagonistic interactions, representing similar and opposing perspectives on the severity of the epidemic, respectively.
We propose an Opinion-Dependent 
Reproduction Number to characterize the mutual influence between epidemic spreading and opinion dissemination over the networks. Through stability analysis of the equilibria, 
we explore the impact of opinions on both epidemic outbreak and eradication, characterized by bounds on the Opinion-Dependent 
Reproduction Number.
We also show how to eradicate 
epidemics by reshaping the opinions, offering researchers 
an approach for designing
control strategies 
to
reach
target audiences to ensure
effective epidemic suppression.
\end{abstract}




%
\IEEEpeerreviewmaketitle

\section{Introduction}
\subsection{Motivation}
Epidemiological models have been extensively studied 
for the purpose of understanding the spread of infectious diseases through  
societies \cite{kermack1927_sir, mei2017epidemics_review, nowzari2016epidemics, pare2020modeling_review}. 
A key epidemiologic metric in these models is 
the basic reproduction number $\left(R_{0}\right)$, 
which 
describes the expected number of cases directly generated by one case in an infection-free population; similarly, the effective reproduction number $\left(R_{t}\right)$ characterizes the average number of new infections caused by a single infected individual at time $t$ in the partially susceptible population\cite{nishiura2009effective}.  
These reproduction numbers can be affected by a variety of factors, including biological properties 
of the epidemic, environmental conditions, and the opinions and behaviors of the 
population \cite{cinelli2020covid}. 
Communities believing that 
an outbreak is
severe will react quickly to suppress the epidemic, in terms of policy-making, rule-following, etc\cite{chan2020social}. On the other hand, misinformation or disbelief 
in an epidemic
could potentially lead 
communities to react incorrectly, perpetuating outbreaks~\cite{kouzy2020coronavirus}. 
\bshe{Further, 
social media significantly increases the rate at which opinions and misinformation spread through communities. During the ongoing SARS-CoV-2 pandemic, the adoption of masks, social-distancing, and vaccination has been significantly influenced by highly polarized attitudes about the virus, as well as misinformation\cite{czeisler2020public}. The polarized opinions towards the seriousness of the epidemic shift
people's reactions towards 
the virus
in different ways, leading to drastic differences in infections over different communities/areas \cite{yeung2020face}.
Given the significant 
link between opinion polarization and pandemic-related outcomes, it is of great importance to provide a foundational understanding of epidemic spreading processes under the effects of 
polarized opinions.}
Motivated by the critical role that opinion dynamics plays in the epidemic spreading (and vice versa), in this work we study a
networked epidemic model coupled with 
a
networked opinion model possessing both cooperative and antagonistic 
interactions 
to understand the mutual influence between epidemic spreading and opinion dissemination over communities. Below, we discuss related literature 
and then describe our contributions. 

\subsection{Literature Review}

\bshe{Researchers have developed and studied threshold conditions and equilibria of networked SIS epidemic models to characterize 
virus spreading over communities \cite{fall2007SIS_model}, as well as the connections between the networked SIS epidemic model and Markov chain models \cite{OmicTN09}}.
Networked epidemic models have also attracted considerable attention
\cite{mei2017epidemics_review, nowzari2016epidemics, pare2020modeling_review,khanafer2016SIS_positivesystems} for modeling the spread of malware in computer networks \cite{berger2005spread} and attacks in cyber-physical systems \cite{soheil}. 
\bshe{On the opinions side, the Altafini model \cite{altafini2013antagonistic_interactions} lays a foundation for various extensions on modeling and control of opinions spreading on signed networks, capturing both cooperative and antagonistic interactions. Therefore, }
to capture the co-existence of 
rival opinions among different populations, we \bshe{follow the Altafini model by using} networks with both positive and negative edges to characterize the cooperative and antagonistic interactions, respectively, of opinion exchange dynamics between communities, as studied in \cite{altafini2013antagonistic_interactions,proskurnikov2015opinion,shi2019dynamics}. 
\bshe{Recent work has combined disease spreading models with human awareness models \cite{hu2018epidemic,granell2014competing,funk2009spread,Paarporn_tcss,pare2020multi_human},} where the infection rates scale with the human awareness toward the epidemic. However, these models lack an explicit dynamical model to represent the change of the population's perception on the severity of the epidemics over time\cite{ruf2019going,panja2020opinion,weihao2020_opinion}.

To 
couple the networked
SIS 
model with opinion dynamics, we employ the health belief model, which is 
the best known and most widely used theory in health behavior research~\cite{healthbelief}.
%
The health belief model proposes that people's beliefs\footnote{ In this article, beliefs, attitudes, and opinions are used interchangeably.} about health problems, perceived benefits of actions, and/or perceived barriers to actions can explain their engagement, or lack thereof, in health-promoting behavior. Therefore, people's beliefs in their perceived susceptibility and/or in their perceived severity of the illness affect how susceptible they are and/or how effective they will be at healing from these epidemics. 
In this article, we 
model the population's beliefs of the severity of the epidemic using opinion dynamics with both cooperative \textit{and} antagonistic interactions. 


\subsection{Contribution}
The contributions of this work are the following.
\begin{enumerate}
\item We develop a networked SIS model coupled with both cooperative and antagonistic opinion dynamics. The opinion dynamics evolves on an opinion-dependent sign switching topology, which characterizes the change of the opinions of the communities toward the epidemic over time.
\item We define an Opinion-Dependent 
Reproduction Number $\left(R_t^{o}\right)$ of the epidemic model. We use $R_t^{o}$ 
to characterize the severity of the epidemic (mild, moderate, severe), equilibria, and stability, which reveal the mutual influence between epidemic spreading and opinion dissemination over the communities.
\item We interpret our results in the context of real-world phenomena under the SARS-CoV-2 pandemic. We propose ways to guide control design and select target communities to shift the opinions of the communities in order to better control the epidemic.
\end{enumerate}

\bsheup{Note that compared to the previous work \cite{weihao2020_opinion} which considers a networked SIS epidemic model with cooperative opinion dynamics, in order to characterize more realistic interactions in opinion formation, we propose an opinion-dependent sign switching structure on opinion dynamics in this work. In particular, we consider opinion dynamics with both cooperative and antagonistic interactions, where the category (either cooperative or antagonistic interaction) of the interactions is determined by similar/polarized opinions toward the seriousness of the epidemic. The switching opinion dynamics with both cooperative and antagonistic interactions leads to 
diverse behavior of opinion formation during the epidemic. 
Unlike \cite{weihao2020_opinion}, we define an Opinion-Dependent Effective Reproduction Number to characterize different epidemic/opinion behavior under multiple virus settings (mild, moderate, severe virus), to capture the coupled behavior of opinion dynamics and epidemic spreading under different situations. 
In particular, we propose strategies to eradicate the epidemic through reshaping the opinions over communities, which is not considered in \cite{weihao2020_opinion}.}

\subsection{Outline of the article}
This work is organized as follows. In Section II, we state the motivation and present the Epidemic-Opinion network model. 
Section III introduces the preliminaries that are used throughout this work. 
In Section IV, we study  
the equilibria of the Epidemic-Opinion network model. Section~V defines the Opinion-Dependent Reproduction Number, which characterizes 
the behavior of the epidemic spreading process. Section V also explores how to  
influence the opinions in order to suppress the outbreak.
Section VI and Section VII present simulation and conclusion/future works, respectively. \bshe{Note that all proofs are included in the Appendix.}

\subsection{Notations}
For any positive integer $n$, we use $[n]$ to denote the index set
$\left\{ 1,2,\ldots,n\right\} $. We view vectors as column vectors
and write $x^{T}$ to denote the transpose of a column vector $x$. We use $x^{n}$ to denote the vector, of the same size as $x$, whose each entry equals the $n$th power of the corresponding entry of $x$. For a
vector $x$, we use $x_{i}$ to denote the $i$th entry.
For any matrix $M\in\R^{n\times n}$, we use $M_{i,:}$, $M_{:,j}$,
$M_{ij}$ ($[M]_{ij}$), to denote its $i$th row, $j$th column, and $ij$th
entry, respectively. We use $M=\diag\left\{ m_{1},\ldots,m_{n}\right\} $
to represent a diagonal matrix $M\in\R^{n\times n}$ with
$M_{ii}=m_{i}$, $\forall i\in\left[n\right]$. We use $\boldsymbol{0}$
and $\boldsymbol{e}$ to denote the vectors whose entries all equal 0 and 1,
respectively, and $I$ to denote the identity matrix. The dimensions
of the vectors and matrices are to be understood from the context.
Let $\partial\left[c,d\right]^{n}$ and $\mathrm{Int}\left[c,d\right]^{n}$ denote
the boundary and interior of the cube $\left[c,d\right]^{n},$ $c,d\in \R$, respectively.

For a real square matrix $M$, we use $\rho\left(M\right)$ and $s\left(M\right)$ to denote
its spectral radius and spectral abscissa (the largest real part among its eigenvalues), respectively. For any two vectors $v,w\in\R^{n}$,
we write $v\geq w$ if $v_{i}\geq w_{i}$, 
and $v\gg w$ if $v_{i}>w_{i}$, $\forall i\in\left[n\right]$.
The comparison notations between vectors are applicable for matrices as well, 
for instance, for $A,B\in\R^{n\times n}$, $A\gg B$ indicates
that $A_{ij}>B_{ij}$, $\forall i,j\in\left[n\right]$. For any
two sets $A$ and $B$, we use $A\setminus B$ to denote the set of
elements in $A$ but not in $B$.
We also employ a modified \bshe{sign} function:
\[
sgnm\left(x\right)=\begin{cases}
\begin{array}{cc}
1, & \mathrm{if}\,\,\,\text{\,\,\ensuremath{x\geq0}}\\
-1, & \mathrm{if}\,\,\,\,\,x<0.
\end{array} & \end{cases}
\]
The Dirac delta function, which is
the first derivative of $\frac{1}{2}sgnm\left(\cdot\right)$, is represented by $\theta\left(\cdot\right)$. \bshe{ Note that we use $sign(\cdot)$ to represent the original sign function, where $sign(x)=sgnm(x)$, $\forall x\neq 0$, and $sign(0)=0$. We use the modified sign function $sgnm(\cdot)$ to classify non-negative and negative opinion states, and use the original sign function $sign(\cdot)$ to distinguish between positive and negative edge weights.}
%

Consider a directed graph $G=\left(\mathcal{V},\mathcal{E}\right)$, with the node set $\mathcal{V}=\left\{ v_{1},\ldots,v_{n}\right\} $
and the edge set $\mathcal{E}\subseteq \mathcal{V}\times \mathcal{V}$. Let
matrix $A=\left[a_{ij}\right]\in\R^{n\times n}$ denote the adjacency matrix of $G=\left(\mathcal{V},\mathcal{E}\right)$, where $a_{ij}\in\R$
if $\left(v_{j},v_{i}\right)\in \mathcal{E}$ and $a_{ij}=0$ otherwise.
Graph $G$ does not allow self-loops, i.e., $a_{ii}=0,$ $\forall i\in \left[n\right]$. Let
$k_{i}=\sum_{j\in \mathcal{N}_{i}}\left|a_{ij}\right|$, where $\mathcal{N}_{i}=\left\{ \left.v_{j}\right\vert \left(v_{j},v_{i}\right)\in \mathcal{E}\right\}$
denotes the neighbor set of $v_{i}$ and $\left|a_{ij}\right|$ denotes
the absolute value of $a_{ij}$. The graph Laplacian of $G$ is defined
as $L\triangleq K-A$, where $K\triangleq \diag\left\{ k_{1},\ldots,k_{n}\right\} $ \bshe{ denotes the degree matrix of the graph $G$.}


\section{Modeling and Problem Formulation} \label{sec:model}

In this section, we introduce an SIS model, coupled with an opinion dynamics model. In particular, we assume
an epidemic 
is
spreading over a group of communities, where the interactions between the communities 
facilitate
the spreading. Furthermore, 
the
opinion of each community 
about
the epidemic 
evolves as a function of the community's infected proportion, the community's own opinion, 
and the opinions of other communities. 

\subsection{Epidemic Dynamics}
Consider an epidemic spreading over $n$ connected communities represented
by a directed graph $G=\left(\mathcal{V},\mathcal{E}\right)$, where the node set $\mathcal{V}=\left\{ v_{1},\ldots,v_{n}\right\}$ 
and the edge set $\mathcal{E}\subseteq \mathcal{V}\times \mathcal{V}$ 
represent the communities
and the epidemic spreading interactions, respectively. The epidemic
spreading interactions over $n$ communities are captured by an adjacency
matrix $A=\left[a_{ij}\right]\in\R^{n\times n}$, 
where $a_{ij}\in \R_{\geq 0}$. 
A directed edge $\left(v_{j},v_{i}\right)$
indicates that community $j$ can
infect community $i$.


We use a networked SIS model to capture the epidemic dynamics 
of the $n$ communities, such that $x_{i}\in\left[0,1\right]$ represents
the proportion of the infected population in 
community $i$, $i\in \left[n\right]$. Note that $x_{i}\times100\%\in\left[0\%,100\%\right]$. In this work, $x_i$ and $x_i\left(t\right)$ are used interchangeably.
\bshe{Although the behavior of an epidemic spreading over communities can be complex \cite{stegehuis2016epidemic}, we leverage one of the epidemic dynamics proposed in \cite{fall2007SIS_model} to capture the evolution of each of the ${n}$ communities}:
\begin{equation}
\dot{x}_{i}\left(t\right)=-\delta_{i}x_{i}\left(t\right)+\left(1-x_{i}\left(t\right)\right)\sum_{j\in \mathcal{N}_{i}}\beta_i a_{ij}x_{j}\left(t\right),\label{eq:ep_1}
\end{equation} 
where $\delta_{i}\in\R_{\geq 0}$ is the average curing rate of community $i$, and $\beta_i a_{ij}\in\R_{\geq 0}$ is the average infection rate of community $j$
to community~$i$.

\subsection{Opinion Dynamics}\label{sec:model_o}
We let the opinions disseminate 
over a graph $\bar{G}=\left(\mathcal{V},\mathcal{\bar{E}}\right)$, 
with the adjacency matrix $\bar{A}=\left[\bar{a}_{ij}\right]\in\R^{n\times n}$, $\bar{a}_{ij}\in \R$.
The opinion 
graph $\bar{G}$ allows both positive and negative edge weights, 
representing cooperative and antagonistic interactions on opinions
exchanged between the communities, respectively. Let $\bar{A}_{u}\in\R^{n\times n}$, with $\left[\bar{A}_{u}\right]_{ij}\in\R_{\geq0}$, and 
$\left[\bar{A}_{u}\right]_{ij}=\left|\bar{a}_{ij}\right|$. Therefore, $\bar{A}_{u}$ captures the coupling strength  of the opinion dynamics without considering the signs.
Similar to \bshe{the definition of graph $G$},
we define the neighbor set of node $v_i$ in graph $\bar{G}$ as $\mathcal{\bar{N}}_i$, and use
$\bar{K}$ and $\bar{L}$ to denote the degree matrix and Laplacian
matrix of $\bar{G}$, respectively. 

For all $t\geq0$, $o\left(t\right)\in \R^{n}$ is the
opinion vector of the $n$ communities,
$o_{i}\left(t\right)\in\left[-0.5,0.5\right]$, $i\in\left[n\right]$. \baike{The range $o_{i}\left(t\right)\in\left[-0.5,0.5\right]$
denotes the belief of 
community $i$ about the severity of the epidemic at time $t$. Note that $o_i\left(t\right)$ and $o_i$ are used interchangeably in this work. 
The opinion $o_{i}\left(t\right)=0.5$ indicates that community $i$ considers the epidemic to be extremely serious, while $o_{i}\left(t\right)=-0.5$ implies that community $i$ thinks the epidemic is not worth addressing. We assume that communities with a neutral opinion $o_{i}\left(t\right)=0$ marginally lean toward treating the epidemic as a threat.}

\phil{
It is natural
to consider that communities with the same attitude toward the epidemic
exchange their opinions cooperatively, while communities with different
attitudes exchange their opinions antagonistically. 
Therefore, b}ased on the communities'
beliefs toward the epidemic at any given time, 
\phil{we allow the edge signs of the opinion graph $\bar{G}$ to switch. We achieve this behavior by} 
partitioning the node set  \phil{of the communities} $\mathcal{V}$ into two
groups, \bs{$\mathcal{V}_{1}\left(o(t)\right)=\left\{ v_{i}\in\mathcal{V}\mid sgnm\left(o_{i}\left(t\right)\right)=1, i\in\left[n\right]\right\}$ and $\mathcal{V}_{2}\left(o(t)\right)=\left\{ v_{i}\in\mathcal{V}\mid sgnm\left(o_{i}\left(t\right)\right)=-1, i\in\left[n\right]\right\} $.}
\phil{Then}
we construct
the adjacency matrix $\bar{A}$ of graph $\bar{G}$ as $\bar{A}\left(o\left(t\right)\right)=\Phi\left(o\left(t\right)\right)\bar{A_{u}}\Phi\left(o\left(t\right)\right)$,
where 
$\Phi\left(o\left(t\right)\right)=\diag\{sgnm\left(o_{1}(t)\right),\ldots,sgnm\left(o_{n}(t)\right)\}$ is a gauge transformation matrix. The entries of the gauge transformation matrix $\Phi\left(o\left(t\right)\right)$ are chosen as $\Phi_{ii}\left(o_{i}\left(t\right)\right)=1$
if \bs{$i\in \mathcal{V}_{p}\left(t\right)$ and $\Phi_{ii}\left(o_{i}\left(t\right)\right)=-1$ if $i\in \mathcal{V}_{q}\left(t\right)$, $p\neq q$,
and $p,q\in\{1,2\}$, $\forall i\in\left[n\right]$.} Through the construction of $\bar{A}\left(o\left(t\right)\right)$, 
$\left|\bar{a}_{ij}\right|$ is fixed and $sign\left(\bar{a}_{ij}\right)$
is switchable. In particular, for all non-zero entries in $\bar{A}\left(o\left(t\right)\right)$, $sign\left(\bar{a}_{ij}\right)=1$ if
$sgnm\left(o_{j}\left(t\right)\right)$ and $sgnm\left(o_{i}\left(t\right)\right)$ are
the same, otherwise, $sign\left(\bar{a}_{ij}\right)=-1$. Therefore,
$\bar{A}\left(o\left(t\right)\right)$ captures the switch of the
opinion interactions through the attitude changing toward the epidemic. Note that an opinion graph $\bar{G}$ constructed 
in this manner
is always structurally balanced by the following definition and lemma.
%
\begin{definition}{\label{SB}} [Structural Balance \cite{altafini2013antagonistic_interactions}]
\textcolor{black}{\label{def:SB} A signed graph $\bar{G}=\left(V,\bar{E}\right)$
is structurally balanced if the node set $\mathcal{V}$ can be partitioned
into $\mathcal{V}_{1}$ and $\mathcal{V}_{2}$ with $\mathcal{V}_{1}\cup\mathcal{V}_{2}=\mathcal{V}$
and $\mathcal{V}_{1}\cap\mathcal{V}_{2}=\emptyset$, where $\bar{a}_{ij}\geq0$
if $v_{i},v_{j}\in\mathcal{V}_{q}$, $q\in\{1,2\},$ and $\bar{a}_{ij}\leq0$
if $v_{i}\in\mathcal{V}_{q}$ and $v_{j}\in\mathcal{V}_{r}$, $q\neq r$,
and $q,r\in\{1,2\}.$}
\end{definition}
\begin{lemma}
\label{Lem:SB} \cite{altafini2013antagonistic_interactions} A connected signed graph
$\bar{G}$ is structurally balanced if and only if there exists a gauge transformation
matrix $\Phi=\diag\left\{ \phi_{1},\ldots,\phi_n\right\} \in\R^{n\times n}$,
with $\phi_{i}\in\left\{ \pm1\right\}$, such that \bs{$\Phi \bar{A}\Phi\in\R^{n\times n}$}
is non-negative.
\end{lemma}


After defining the opinion interaction matrix $\bar{A}\left(o\left(t\right)\right)$,
a variant of the opinion dynamics model in \cite{altafini2013antagonistic_interactions}
evolving over the $n$ communities
with both cooperative and antagonistic interactions 
is given by 
\begin{align}
\dot{o}_{i}\left(t\right)=\sum_{j\in \mathcal{\bar{N}}_{i}}\left|\bar{a}_{ij}\left(o\left(t\right)\right)\right|\left(sign\left(\bar{a}_{ij}\left(o\left(t\right)\right)\right)o_{j}\left(t\right)-o_{i}\left(i\right)\right),\label{eq:op_1}
\end{align}
with the compact form 
\begin{equation}
\dot{o}\left(t\right)=-\Phi\left(o\left(t\right)\right)\bar{L}_{u}\Phi\left(o\left(t\right)\right) o\left(t\right),\label{eq:opi_sb}
\end{equation}
where $\bar{L}_{u}$ represents the Laplacian matrix of $\bar{A}_{u}$. \bshe{An example is shown in the Appendix to illustrate the evolution of opinion switching dynamics in \eqref{eq:opi_sb}.}  

\subsection{Coupled Epidemic-Opinion Dynamics}
Having introducing the networked SIS epidemic model spreading over
the $n$ communities, and opinions spreading over the
same $n$ communities, 
we now introduce network dynamical models that couple
the epidemic dynamics with the opinion dynamics.

Assume that a community's opinion/attitude toward the severity of
an epidemic will affect its actions, which leads
to the variation of the community's average healing rate and infection
rate. For instance, a community being very cautious about the epidemic
will broadcast the influence of the epidemic more frequently, and make policies to suppress the epidemic, 
and the people in that community 
will be
more likely to follow the instructions given by scientific institutions, and seek treatments in a timely manner. These actions will 
result in the community having a
lower average infection rate and a higher average healing rate.
To better describe the situation, we use $\delta_{\min}$
and $\beta_{\min}$ to denote the possible minimum average healing rate and infection rate
for all communities, respectively.\footnote{We assume homogeneous minimum healing and infection rates for simplicity. The results in this work can be extended to the heterogeneous case.}
To incorporate the opinion dynamics
in \eqref{eq:op_1} into the epidemic dynamics in \eqref{eq:ep_1},
we consider
\begin{align}
\dot{x}_{i}\left(t\right) & =-\left[\delta_{\min}+\left(\delta_{i}-\delta_{\min}\right)o'_{i}\left(t\right)\right]x_{i}\left(t\right)\nonumber \\
 & +\left(1-x_{i}\left(t\right)\right)\sum_{j\in \mathcal{N}_{i}}\left[\beta_{ij}-\left(\beta_{ij}-\beta_{\min}\right)o'_{i}\left(t\right)\right]x_{j}\left(t\right),\label{eq:epi}
\end{align}
where $o'_{i}\left(t\right)=o_{i}\left(t\right)+0.5$ shifts the opinion into the range $\left[0,1\right]$. In particular, the term $o'_{i}\left(t\right)$ scales the average healing and infection rates between their maximum and minimum values. 
In the case when $o_{i}\left(t\right)=-0.5$, which
implies that community $i$ does not consider the epidemic 
a threat at time $t$,
community $i$ will take no action to protect itself and thus is maximally
exposed to the infection. In the case when $o_{i}\left(t\right)=0.5$, which implies
that community $i$ believes the epidemic is extremely serious, it will
implement 
policies and limitations to decrease the infection rate and
seek out all possible medical treatment options 
to improve its
healing rate. Therefore, the model allows the communities' opinions to
affect how susceptible they are and how effectively they heal from
the virus, capturing the health belief model \cite{healthbelief}, as explained  
in the Introduction.

The proportion of the infected population of a community~$i$ can also have an effect on its opinion/attitude toward the epidemic. 
Consider the epidemic dynamics in \eqref{eq:ep_1} incorporated
with \eqref{eq:op_1} as follows:
\begin{align}
\dot{o}_i(t) & =\left(x_{i}\left(t\right)-o'_{i}\left(t\right)\right)\nonumber \\
 & \text{}+\sum_{j\in \mathcal{\bar{N}}_{i}}\left|\bar{a}_{ij}(o(t))\right|\left(sign\left(\bar{a}_{ij}(o(t))\right)o_{j}\left(t\right)-o_{i}\left(t\right)\right)\label{eq:opi}.
\end{align}
The first term on the right hand side of \eqref{eq:opi} captures how the infected proportion of a community 
affects
its own opinion. If \bshe{$o'_{i}\left(t\right)$} is
small but the community is heavily infected, i.e., $x_{i}\left(t\right)$ is large, \bshe{$o'_{i}\left(t\right)$} will increase.
If \bshe{$o'_{i}\left(t\right)$} is large but the community has few infections, i.e., $x_{i}\left(t\right)$ is small, \bshe{$o'_{i}\left(t\right)$}
will decrease. This behavior is sensible since a community's infection
level should affect its belief 
in the severeness of the virus,
which is consistent
with the health belief model in \cite{healthbelief}. 
The second term on the right hand side of \eqref{eq:opi} is from \eqref{eq:op_1}. The neighbors of 
community~$i$ affect its opinion cooperatively ($sign\left(\bar{a}_{ij}(o(t))\right)=1$)
or
antagonistically ($sign\left(\bar{a}_{ij}(o(t))\right)=-1$).

\subsection{Problem Statements}
\baike{Now that we have presented the 
Epidemic-Opinion
model in \eqref{eq:epi} and \eqref{eq:opi}, we can state the problem of interest in this work. We are interested in exploring the mutual influence between the epidemic spreading over $n$ communities captured by graph $G$ in \eqref{eq:epi} 
and the opinions of the $n$ communities about the epidemic captured by graph $\bar{G}$ in \eqref{eq:opi}. 
We will analyze 
the equilibria of the system in \eqref{eq:epi}-\eqref{eq:opi}
under different settings. In particular, we will define an Opinion-Dependent Reproduction Number to characterize the spreading of the virus. 
We will study the stability of the equilibria of the system to infer the behaviors of the epidemic and opinions spreading over the $n$ communities. Finally, 
we will generate strategies to eradicate the epidemic by affecting the opinion states of the system, which could potentially guide the use of social media to  
broadcast the severity of the pandemic to appropriate communities to suppress the epidemic.}

\section{Preliminaries}
We impose the following natural restrictions on the parameters of
the models throughout the article.

\begin{assumption}
\label{A1}
\baike{
Let
$x_{i}\left(0\right)\in\left[0,1\right]$,  $o_{i}\left(0\right)\in\left[-0.5, 0.5\right]$,
$\delta_{i}\ge\delta_{{\rm min}}>0$, and $\beta_{ij}\ge\beta_{{\rm min}}>0$, $\forall i\in\left[n\right]$ and $\forall j\in \mathcal{N}_{i}$. }The epidemic spreading graph $G$ and opinion
dissemination graph $\bar{G}$ are strongly connected.
\end{assumption}

Note that the adjacency matrix of a strongly connected graph is irreducible. Therefore, the adjacency matrix $A$ of the graph $G$ is a nonnegative irreducible matrix. A real square matrix $M$ is called
a Metzler matrix if $M_{ij}\geq0$, $\forall i,j\in\left[n\right]$ \baike{and $i\neq j$}, which \baike{implies}
that the adjacency matrix $A$ is \baike{also} an irreducible Metzler matrix. Some of our results rely on properties of Metzler matrices and nonnegative matrices, which
we briefly recall below.

\begin{lemma}
\cite [Prop. 2]{rantzer2011distributed}
\label{prop:Metzler}For a Metzler matrix $M\in\R^{n\times n}$, the following statements are equivalent:
\begin{enumerate}
\item The matrix M is Hurwitz;
\item \textcolor{black}{There exists a vector $v\gg\boldsymbol{0}$ such that $Mv\ll\boldsymbol{0}$};
\item \baike{There exists a vector $u\gg\boldsymbol{0}$ such that $u^{T}M\ll\boldsymbol{0}$};
\item \baike{There is a positive diagonal matrix Q such that 
$M^{T}Q+QM$ is negative definite.}
\end{enumerate}
\end{lemma}

\begin{lemma}
\label{lem:semi_d}
\cite [Lemma A.1]{khanafer2016SIS_positivesystems}
For an irreducible Metzler matrix $M\in\R^{n\times n}$,
if $s\left(M\right)=0$, there exists a positive diagonal matrix Q such that 
$M^{T}Q+QM$ is negative semidefinite.
\end{lemma}

\begin{lemma}
\label{prop:irr_non}\cite[Thm. 2.7, and Lemma 2.4]{varga2009matrix_book} Suppose
that M is an irreducible nonnegative matrix. Then, the following statements hold:
\begin{enumerate}
\item M has a \baike{simple} positive real eigenvalue equal to its spectral radius, $\rho(M)$;
\item There is \baike{a unique (up to scalar multiple)} eigenvector $v\gg\boldsymbol{0}$ corresponding to $\rho(M)$;
\item $\rho(M)$ increases when any entry of M increases.
\end{enumerate}
\end{lemma}

\begin{lemma}
\baike{\cite[Sec. 2.1 and Lemma 2.3]{varga2009matrix_book}} 
\label{lem:irr_M} Suppose that $M$ is an irreducible Metzler matrix. Then, $s\left(M\right)$ is a simple eigenvalue of $M$ and
there exists a unique (up to scalar multiple) vector $x\gg \boldsymbol{0}$ such
that $Mx=s\left(M\right)x$. Let $z>\boldsymbol{0}$ be a vector in $\R^{n}$. If
$Mz<\lambda z$, then $s(M)<\lambda$. If $Mz=\lambda z$, then $s(M)=\lambda$.
If $Mz>\lambda z$, then $s(M)>\lambda$.
\end{lemma}

\begin{lemma}
\label{lem:irr_spe}\cite[Prop. 1]{bivirus}
Suppose that $\varLambda$ is a negative diagonal matrix in $\R^{n\times n}$
and $N$ is an irreducible nonnegative matrix in $\R^{n\times n}$.
Let $M=\varLambda+N$. Then, $s(M)<0$ if and only if $\rho(-\varLambda^{-1}N)<1$,
$s(M)=0$ if and only if $\rho(-\varLambda^{-1}N)=0$, and $s(M)>0$
if and only if $\rho(-\varLambda^{-1}N)~>~1$. 
\end{lemma}



\section{ Equilibria  of  Epidemic-Opinion  Dynamics}

This section considers the mutual influence between the epidemic dynamics
in \eqref{eq:epi} and opinion dynamics in \eqref{eq:opi}, and lays 
a foundation for 
analyzing 
the stability and convergence to 
the equilibria of the coupled dynamics under different conditions in the next section.

We write \eqref{eq:epi} and \eqref{eq:opi} in a compact
form as follows:

\begin{align}
\left[\begin{array}{c}
\dot{x}(t)\\
\dot{o}(t)
\end{array}\right] & =\left[\begin{array}{cc}
W(o(t)) & \boldsymbol{0}\\
I & \text{}-\Phi(o(t)) \bar{L}_{u}\Phi(o(t))-I
\end{array}\right]\nonumber \\
 &\ \ \ \ \ \ \ \ \ \ \ \ \ \ \ \ \ \ \text{}\text{}\text{}\times\left[\begin{array}{c}
x(t)\\
o(t)
\end{array}\right]-\left[\begin{array}{c}
\boldsymbol{0}\\
0.5\boldsymbol{e}
\end{array}\right]\label{eq:epi_op},
\end{align}
where
\begin{align*}
W\left(o\left(t\right)\right) & =-\left(D_{\min}+\left(D-D_{\min}\right)\left(O\left(t\right)+0.5I\right)\right)\\
 & +\left(I-X\left(t\right)\right)\left(B-\left(O\left(t\right)+0.5I\right)\left(B-B_{\min}\right)\right),
\end{align*}
$O\left(t\right)=\diag\left\{ o_{1}\left(t\right),\ldots,o_{n}\left(t\right)\right\}$, $X(t)=\diag\left\{ x_{1}(t),\ldots,x_{n}(t)\right\}$,
$D=\diag\left\{ \delta_{1},\ldots,\delta_{n}\right\} $,  $D_{\min}=\delta_{\min}I$, \bs{$B=\left[\beta_{ij}\right]\in\R^{n\times n}$} and $B_{\min}=\beta_{\min}\tilde{A}$, with $\tilde{A}\in\R^{n\times n}$
being the unweighted 
adjacency matrix of graph $G$ \bs{(with $\tilde{A}_{ij}\in\left\{ 0,1\right\}$, $\forall i,j\in\left[n\right]$)}. Note that the Epidemic-Opinion model in \eqref{eq:epi_op} is a nonlinear system, due to the nonlinearity brought by \bs{$W\left(o\left(t\right)\right)$} and $-\Phi\left(o\left(t\right)\right) \bar{L}_{u}\Phi\left(o\left(t\right)\right)$.
The system in \eqref{eq:epi_op} follows:
\begin{equation}
\dot{x}\left(t\right)=-D\left(o\left(t\right)\right)x\left(t\right)+\left(I-X\left(t\right)\right)B\left(o\left(t\right)\right)x\left(t\right)\label{eq:epi_c}
\end{equation}
\begin{equation}
\dot{o}\left(t\right)=Ix\left(t\right)-\left(\Phi\left(o\left(t\right)\right)\bar{L}_{u}\Phi\left(o\left(t\right)\right)+I\right)o\left(t\right)-0.5\boldsymbol{e},\label{eq:op_c}
\end{equation}
where $D\left(o\left(t\right)\right)=D_{\min}+\left(D-D_{\min}\right)\left(O\left(t\right)+0.5I\right)$, and
$B\left(o\left(t\right)\right)=B-\left(O\left(t\right)+0.5I\right)\left(B-B_{\min}\right)$ are the opinion-dependent healing and infection matrices, respectively.


\begin{remark}
\label{re:basics}
Based on Assumption \ref{A1}, $B\left(o\left(t\right)\right)$
is an irreducible non-negative matrix and $D\left(o\left(t\right)\right)$ is
a positive definite diagonal matrix, $\forall t$. 
It can be verified that the matrix $W(o(t))$ is a Metzler matrix. Since graph $\bar{G}$ is strongly connected \baike{and structurally balanced}, the
Laplacian matrix $\ensuremath{\Phi\left(o\left(t\right)\right)\bar{L}_{u}\Phi\left(o\left(t\right)\right)}$ of graph $\bar{G}$ has only one zero
eigenvalue, and the rest \bs{of the} eigenvalues have positive
real parts \cite[\baike{Lemma 2}]{altafini2013antagonistic_interactions}. \baike{Note that the gauge transformation defined in Lemma \ref{Lem:SB} does not change
the spectra of a matrix \cite{altafini2013antagonistic_interactions}, and thus, $\ensuremath{\Phi\left(o\left(t\right)\right)\bar{L}_{u}\Phi\left(o\left(t\right)\right)}$ and
$\bar{L}_{u}$ share the same spectra. Further, the matrix $\left(\ensuremath{\Phi\left(o\left(t\right)\right)\bar{L}_{u}\Phi\left(o\left(t\right)\right)}+I\right)$
has one eigenvalue located at one, and the rest of its eigenvalues have
positive real parts larger than one, which implies that the matrix $-\left(\ensuremath{\Phi\left(o\left(t\right)\right)\bar{L}_{u}\Phi\left(o\left(t\right)\right)}+I\right)$
is Hurwitz, $\forall t$.} Note that the opinion dynamics in \eqref{eq:op_c} is a state-based switching  system. For each subsystem, the matrix $\Phi\left(o\left(t\right)\right)\bar{L}_{u}\Phi\left(o\left(t\right)\right)$ is 
\baike{the} standard signed Laplacian matrix.

\end{remark}

Now that we have introduced 
the Epidemic-Opinion model, to analyze the behavior of the system in \eqref{eq:epi_op}, 
we need to show that the
model is well-defined. 

\begin{lemma}
\label{lem:Inv}For the system defined in \eqref{eq:epi_op}, if $x_{i}\left(t\right)\in\left[0,1\right]$
and $o_{i}\left(t\right)\in\left[-0.5,0.5\right]$, $\forall i\in\left[n\right]$,
then $x_{i}\left(t+\tau\right)\in\left[0,1\right]$ and $o_{i}\left(t+\tau\right)\in\left[-0.5,0.5\right]$,
$\forall$ $i\in\left[n\right]$, and $\forall\tau\geq0$.
\end{lemma}

To explore 
the equilibria of \eqref{eq:epi_op}, let $z\left(t\right)=\left[x^{T}\text{\ensuremath{}}o^{T}\right]^{T}$
denote the states of the system in \eqref{eq:epi_op}, $z\left(t\right)\in\R^{2n}$, and $z^{*}=\left[\left(x^{*}\right)^{T}\left(o^{*}\right)^{T}\right]^{T}$
denote an equilibrium of \eqref{eq:epi_op}. \bs{We say} $x^{*}$ and $o^{*}$ 
are
the 
equilibria of \eqref{eq:epi_c} and \eqref{eq:op_c}, respectively. 
%
\begin{definition}\label{def:eq}
Let the state $\left(x^{*},o^{*}\right)$ denote an equilibrium of
\eqref{eq:epi_op}, 
where $\left(x^{*},o^{*}\right)$ is
\begin{enumerate}
\item \textcolor{black}{a consensus-healthy state if }$\left(x^{*},o^{*}\right)=\left(\boldsymbol{0},o^{*}\right),$ and
$o_{i}^{*}=o_{j}^{*}$, $\forall i,j\in\left[n\right]$;
\item \textcolor{black}{a dissensus-healthy state if }$\left(x^{*},o^{*}\right)=\left(\boldsymbol{0},o^{*}\right)$,
and $\exists i,j\in\left[n\right]$, s.t. $o_{i}^{*}\neq o_{j}^{*}$;
\item a \textcolor{black}{consensus-endemic state if }
\baike{$x^{*}\geq\boldsymbol{0}$}, $x^{*}\neq\boldsymbol{0}$,
and $o_{i}^{*}=o_{j}^{*}$, $\forall i,j\in\left[n\right]$;
\item a \textcolor{black}{dissensus-endemic state if }
\baike{$x^{*}\geq\boldsymbol{0}$}, $x^{*}\neq\boldsymbol{0}$,
and $\exists i,j\in\left[n\right]$, s.t. $o_{i}^{*}\neq o_{j}^{*}$\phil{.} 
\end{enumerate}
\end{definition}
\baike{In this work, we use the term \textit{healthy state} to describe both case 1) and case 2) in Definition \ref{def:eq}, and the term \textit{endemic state} 
for
both case 3) and case 4)}. It is obvious that the Epidemic-Opinion model in \eqref{eq:epi_op}
has a consensus-healthy state $\left(\boldsymbol{0},-0.5\boldsymbol{e}\right)$ as its trivial equilibrium. Further, from \eqref{eq:op_c}, when $o^{*}=-0.5\boldsymbol{e}$, 
\begin{align*}
\boldsymbol{0} & =x^{*}+\left(\bar{L}_{u}+I\right)\times0.5\boldsymbol{e}-0.5\boldsymbol{e}\\
 & =x^{*}+\bar{L}_{u}\times0.5\boldsymbol{e}+0.5\boldsymbol{e}-0.5\boldsymbol{e}\\
 & =x^{*},
\end{align*}
which indicates $x^{*}=\boldsymbol{0}$. Therefore, the consensus state $o^{*}=-0.5\boldsymbol{e}$ must pair with the healthy state $x^{*}=\boldsymbol{0}$.
The following theorem summarizes 
the healthy equilibria of \eqref{eq:epi_op}.

\begin{theorem}
\label{thm:equi_basic}For the Epidemic-Opinion model in \eqref{eq:epi_op}, a healthy equilibrium 
$z^{*}$
is either the 
\baike{unique}
consensus-healthy state $\left(x^{*}=\boldsymbol{0}, o^{*}=-0.5\boldsymbol{e}\right)$, or a dissensus-healthy state $\left(x^{*}=\boldsymbol{0}, o^{*}\right)$, with $o^{*}=\left(\Phi\left(o^{*}\right) \bar{L}_{u}\Phi\left(o^{*}\right)+I\right)^{-1}\left(-0.5\boldsymbol{e}\right)$ with both positive and negative entries, and $o^{*}_i\in[-0.5,0.5],\forall i\in [n]$.
\end{theorem}
Theorem \ref{thm:equi_basic} states that the dissensus opinion equilibrium
$o^{*}=(\Phi(o^{*}) \bar{L}_{u}\Phi(o^{*})+I)^{-1}(-0.5\boldsymbol{e})$ is always in the set $[-0.5,0.5]^n$, i.e., its value is always consistent with its physical meaning.


\begin{corollary}
\label{cor:unique}
For the Epidemic-Opinion model in \eqref{eq:epi_op}, 
the consensus-healthy state $\left(x^{*}=\boldsymbol{0}, o^{*}=-0.5\boldsymbol{e}\right)$ is the unique equilibrium with consensus in opinions, that is, $o^{*}=\alpha\boldsymbol{e}$, $\alpha\in\left[-0.5,0.5\right]$.
\end{corollary}



Note that if \bs{$\exists i\in\left[n\right]$, s.t. $o^{*}_i=0$,} from \eqref{eq:op_1}, it must be true that $o^{*}_i=0$, $\forall i\in\left[n\right]$. However, Corollary \ref{cor:unique} states that $z^{*}$ does not include the case that $o^{*}=\boldsymbol{0}$. Therefore, we have the next corollary.

\begin{corollary}
\label{cor:not_0}
For the Epidemic-Opinion model in \eqref{eq:epi_op}, the equilibrium $z^{*}$ cannot include $o^{*}_i=0$, $\forall i\in\left[n\right]$.
\end{corollary}

\begin{remark}
Theorem \ref{thm:equi_basic} indicates 
the possible
equilibria
when the epidemic dies out. The consensus-healthy state $\left(x^{*}=\boldsymbol{0}, o^{*}=-0.5\boldsymbol{e}\right)$ implies that, at the stage that the epidemic disappears, all communities agree that the epidemic is not a threat. However, the existence of dissensus-healthy states $\left(x^{*}=\boldsymbol{0}, o^{*}\right)$, with $o^{*}$ having both positive and negative entries, describes
the scenario
when communities hold different beliefs toward the epidemic 
at
the time
when
the epidemic is about to disappear, 
and thus
the epidemic will still 
cause
possible contention
between different communities. Corollary \ref{cor:unique} states that the only way to ensure all communities 
reach agreement is that they all agree that the epidemic is not a threat at the moment the epidemic dies out. 
In other words, it is impossible for all communities to agree that the epidemic is not worth treating seriously while the epidemic is still spreading. \bshe{Note that if there are no antagonistic interactions, following a similar procedure as in the proof of Theorem 1, one can prove that all communities will always reach the consensus point corresponding to the belief that the epidemic is not serious when the epidemic disappears.}
\end{remark}

After analyzing the healthy equilibria, the following lemma further explores  the endemic equilibria of \eqref{eq:epi_op}. 

\begin{lemma}
\label{lem:equi_non}If $\left(x^{*},o^{*}\right)$ is an endemic equilibrium
of the system in $\eqref{eq:epi_op}$, 
then $\boldsymbol{0}\ll x^{*}\ll\boldsymbol{e}$, $-0.5\boldsymbol{e}\ll o^{*}\ll0.5\boldsymbol{e}$.
\end{lemma}

\begin{remark}
Lemma \ref{lem:equi_non} states that
if an endemic state
exists, no community can be
completely infection free or completely infected.
Further, the
equilibrium
of a community cannot be equal to
one of the extreme
beliefs.
\end{remark}

\section{Stability Analysis of 
Epidemic-Opinion  Dynamics}

In this section, we analyze the properties of the equilibria of the Epidemic-Opinion \baike{model} in \eqref{eq:epi_op}, to reveal the mutual influence between disease spreading and opinion formation
during an epidemic.

As mentioned in \cite{nishiura2009effective}, the 
reproduction number $R$ of an epidemic is critical in determining the spreading of the epidemic. In line with the expression on the 
reproduction number $R$, 
we define an \baike{Opinion-Dependent 
Reproduction Number} $R_{t}^{o}$ to characterize the performance of the Epidemic-Opinion model in \eqref{eq:epi_op}.

\begin{definition}{[Opinion-Dependent 
Reproduction Number] 
Let $R_{t}^{o}=\rho\left(D\left(o\left(t\right)\right)^{-1}B\left(o\left(t\right)\right)\right)$ denote the Opinion-Dependent 
Reproduction Number, where $D\left(o\left(t\right)\right)$ and $B\left(o\left(t\right)\right)$ are defined in \eqref{eq:epi_c}.
\label{def:Rt} }
\end{definition}

Note that the Opinion-Dependent 
Reproduction Number $R_{t}^{o}$ depends on the variation of the opinion states \bs{$o\left(t\right)$}. 
When all communities think the epidemic is extremely serious, by defining $o_{\max}=0.5\boldsymbol{e}$, we have 
\bshe{
\begin{align*}
    R_{\min}=\rho\left(D\left(o_{\max}\right)^{-1}B\left(o_{\max}\right)\right)=\rho\left(D^{-1}B_{\min}\right).
\end{align*}}
Instead, when all communities believe that the epidemic is not real, by defining $o_{\min}=-0.5\boldsymbol{e}$, we have
\bshe{
\begin{align*}
  R_{\max}=\rho\left(D\left(o_{\min}\right)^{-1}B\left(o_{\min}\right)\right)=\rho\left(D_{\min}^{-1}B\right).  
\end{align*}}
\begin{proposition}
\label{prop:Spe_r}The Opinion-Dependent 
Reproduction Number $R_{t}^{o}$
\bs{has} the following 
\bs{properties}:
\begin{enumerate}
\item If $o\left(t_{0}\right)\leq o\left(t_{1}\right)$, then\textcolor{black}{{}
$R_{t_{0}}^{o}\geq R_{t_{1}}^{o}$, and vice versa;}
\item \baike{ $R_{\min}
\leq R_{t}^{o}\leq
R_{\max}$. }
\end{enumerate}
\end{proposition}

\begin{remark}
\bshe{
Recall from the Introduction that reproduction numbers of epidemic models are important metrics to capture the seriousness of epidemics. The reproduction number of an SIS model is determined by 
transmission and healing rates. In reality, 
multiple factors such as social distancing and vaccination have an impact on transmission and healing rates. Compared to previous works that study time-invariant SIS network models \cite{fall2007SIS_model}, and time-varying SIS network models \cite{rami2013stability}, Definition \ref{def:Rt} and Proposition \ref{prop:Spe_r} analyze the reproduction number based on the influence of opinions on transmission and healing rates.}
\end{remark}
Proposition \ref{prop:Spe_r} indicates that the opinions toward the epidemic provide bounds on the Opinion-Dependent 
Reproduction 
Number $R_{t}^{o}$. The more seriously 
community $i$ treats the epidemic, i.e., with a higher $o\left(t\right)$, the lower $R_{t}^{o}$ is, and vice versa. In the following sections, we interpret 
$R_{t}^{o}$ as the severity of an epidemic, 
and explore the behavior of \eqref{eq:epi_op} through 
bounds on 
$R_{t}^{o}$.

\subsection{Mild 
Viruses}
In this section we explore the behavior of viruses that are only slightly contagious, that is,  where $R_{t}^{o}\leq R_{\max}\leq 1$. 
First we analyze the equilibrium of the system in \eqref{eq:epi_op}
under the condition that 
\baike{$R_{\max}\leq 1$}.
\begin{proposition}
\label{prop:case1}If 
\baike{$R_{\max}\leq 1$}, \baike{then every equilibrium of (\ref{eq:epi_op}) is a healthy state.}
\end{proposition}

Proposition \ref{prop:case1} claims that when $R_{\max}\le1$, i.e., the severity of the epidemic is mild, the Epidemic-Opinion model in \eqref{eq:epi_op} has only healthy equilibria, despite 
the evolution of the opinions. To further explore the behavior of 
\eqref{eq:epi_op}, \bshe{we study the stability of the healthy equilibria.}

\begin{proposition}
\label{prop:unstable}If $R_{\max}< 1$, then
all the healthy equilibria $\left(\boldsymbol{0},o^{*}\right)$ of \eqref{eq:epi_op} are locally exponentially
stable. 

\end{proposition}

\begin{theorem}
\label{thm: stability_case1}
If $R_{\max}\leq 1$, \baike{then for any initial condition, the system in \eqref{eq:epi_op} will asymptotically converge to 
\bs{a} healthy equilibrium.
If $R_{\max}<1$, the convergence is exponentially fast.}
\end{theorem}

\begin{remark}
Proposition \ref{prop:unstable} and Theorem \ref{thm: stability_case1} reveal that, under the condition that $R_{\max}\leq 1$, the initial conditions of the epidemic (i.e., the level of the infection in each community) and/or the opinion states (i.e., how much the communities underestimate the severity of the epidemic), will not hinder the epidemic from disappearing quickly. Hence, it is unnecessary to broadcast the severity of the epidemic publicly, since the epidemic will die \bs{out quickly}. 

The initial condition 
may affect the opinions after the epidemic disappears. Imagine the case \bs{where} few infections appear in each community 
\bs{and no community} 
\bs{believes} the epidemic is serious at the beginning. Then, the epidemic will disappear 
\bs{and} all communities \bs{will} reach \bs{a} consensus that the epidemic is not a threat, captured by the unique consensus-healthy equilibrium. 
\bs{Alternatively}, consider the case 
\bs{where} some communities are heavily infected at the beginning, hence they believe the epidemic is a threat. Even after the epidemic dies out quickly, disagreement will 
\bs{linger between} communities, corresponding to the dissensus-healthy equilibria.
\end{remark}
\subsection{
Severe
Viruses}
In this section we explore the behavior of viruses that are very contagious, that is, where $ R_{\min}>1$. Note that Theorem~\ref{thm: stability_case1} demonstrates that the healthy states $\left(\boldsymbol{0}, o^{*}\right)$, are equilibria for \eqref{eq:epi_op} under any $R_{t}^{o}$, s.t. $R_{\max}\leq 1$. Therefore, we consider the stability properties 
of the healthy equilibria.


\begin{lemma}
\label{lem:unstable}Under the condition that $R_{\min}>1$,
all the healthy equilibria $\left(\boldsymbol{0}, o^{*}\right)$ are unstable.
\end{lemma}

\begin{remark}
The proof of Lemma \ref{lem:unstable} follows a similar procedure as in the proof of  Proposition \ref{prop:unstable}, and is thus omitted. Lemma \ref{lem:unstable} states that, when the Opinion-Dependent Reproduction Number is large, $R_{\min}>1$, i.e., the epidemic is highly contagious, even if all communities have zero infection, one infected person appearing in any community will result in an outbreak, leading to a pandemic. Further, with all communities being extremely cautious about the epidemic (\bs{$o^{*}=0.5\boldsymbol{e}$}), taking every action suggested by the scientific institutions, \bs{public health officials}, and the media to protect themselves, the epidemic will 
\bs{continue} spreading. Therefore, relying only on non-pharmaceutical Interventions (NPIs) through social media, \bs{it would be impossible to} eradicate the epidemic. 
\end{remark}

After studying the healthy equilibria in Lemma \ref{lem:unstable}, we explore the existence of the endemic equilibria under the condition that $R_{\min}>1$. Lemma \ref{lem:equi_non} claims that, if \bs{it} exists, the endemic state $\left(x^{*},o^{*}\right)$ must satisfy $\boldsymbol{e}\gg x^{*}\gg\boldsymbol{0}$, $0.5\boldsymbol{e}\gg o^{*}\gg -0.5\boldsymbol{e}$.
Since $\left(-D+B_{\min}\right)$ is an irreducible Metzler matrix, $\rho(D^{-1}B_{\min})>1$
\bs{implies} 
$s(-D+B_{\min})>0$. From Lemma \ref{lem:irr_M}, \bs{let} $\phi\triangleq s(-D+B_{\min})$ \bs{be the} eigenvalue of $\left(-D+B_{\min}\right)$ with an associated right eigenvector $y\gg\boldsymbol{0}$.
Without loss of generality, assume $\underset{i\in[n]}{\max}$ $ y_{i}=1$. Now, define
for any $\epsilon\in[0,1)$, a convex and compact subset of $\chi$ as 
\begin{equation}
\Xi_{\epsilon}\triangleq\{z\in\chi:z_{i}\geq\epsilon y_{i}\,\forall\,i\in[n]\}\label{eq:xi},
\end{equation}
where \begin{align*}
\ensuremath{\chi=\{z\in \R^{2n}\mid z_{i}\in[0,1], i= 1,\ldots, n};\\
z_{i}\in[-0.5,0.5], i= n+1,\ldots, 2n\}.
\end{align*}
Note that $\Xi_{0}=\chi$ and $\forall \epsilon>0$, $\Xi_{\epsilon}\subset\chi$. \baike{From the proof of Lemma \ref{lem:Inv} and the piece-wise continuity of the system in (\ref{eq:epi_op}), we have the following results.} 

\begin{lemma}
\label{lem:set}Consider the system in \eqref{eq:epi_op}. \bshe{If $z(t)\in\partial\chi\setminus\left(\boldsymbol{0}, o^{*}\right)$, where $\left(\boldsymbol{0}, o^{*}\right)$ indicates the set of healthy equilibria of \eqref{eq:epi_op},}
then $z(t+\tau)\in\mathrm{Int}\chi$,  
$\forall \tau\geq 0$. 
\end{lemma}

\begin{theorem}
\label{thm:case2}
Suppose that $R_{\min}>1$. Then, there exists
a sufficiently small $\bar{\epsilon}$ such that $\Xi_{\epsilon}$
defined in \eqref{eq:xi} for every $\epsilon\in(0,\bar{\epsilon}]$
is a positive invariant set for the system in \eqref{eq:epi_op}.
Moreover, $\eqref{eq:epi_op}$ has at least one endemic equilibrium
in $\chi$.
\end{theorem}

Combined with Theorem \ref{thm:equi_basic}, Theorem \ref{thm:case2} states that, when $R_{\min}>1$, the 
system in \eqref{eq:epi_op} \bs{has} both healthy and endemic 
equilibria. Lemma \ref{lem:unstable} shows the healthy equilibria are unstable. \bshe{
Previous work shows the existence of a limit cycle for time-varying networked SIS models under certain conditions \cite{rami2013stability}, where the parameters of the transmission matrix switch. In our current work, the coupling of opinion dynamics changes the parameters of the network  SIS  model, and the opinion network can switch as well. Proving the attractiveness of this set and the lack of limit cycles remains a research direction for future work.}
\subsection{Moderate Viruses}
The previous 
sections show 
\phil{that the opinion states have little impact on the behavior of epidemics} when the epidemic is either highly contagious ($R_{\min}>1$) or very mild ($R_{\max}<1$). 
In this section we explore the behavior of viruses that are 
\bshe{moderately contagious,}
that is, where
$R_{\min}<1$ and $R_{\max}>1$.
\bshe{We show that the moderate viruses cases include properties from both mild and severe viruses, where the behavior of the epidemic is affected by the opinion dynamics. We propose one strategy to analyze  
how the influence of stubborn communities on opinion states 
can
impact the behavior of the epidemic.}
\subsubsection{Healthy State}

\bshe{Recall from Definition \ref{def:Rt} that the Opinion-Dependent Reproduction Number is determined by the opinion states of the system, which are continuous.} Based on Proposition \ref{prop:Spe_r}, $R_{\min}\leq R_{t}^{o}\leq\ R_{\max}$. If $R_{\min}<1$ and $R_{\max}>1$, there must be at least one opinion $o$ such that $R_{t}^{o}=1$. Therefore, the system in \eqref{eq:epi_op} may contain properties that both cases $R_{\min}>1$ and $R_{\max}<1$ have. \bshe{To study eradication strategies of the epidemic, we need to explore the behavior of healthy equilibria}. Since the range of $R_{t}^{o}$ depends on the opinion state $o\left(t\right)$, for healthy equilibria, \bs{we evaluate $R_{t}^{o^{*}}$ 
regarding the opinion state $o^{*}$ at the healthy equilibrium. Thus, we have the following results.}
\begin{theorem}
\label{thm:case_3}
\bs{ For the system in \eqref{eq:epi_op}, if $R_{\min}<1$, and $R_{\max}>1$,} the following statements hold:
\begin{enumerate}
 \item All the dissensus-healthy equilibria  $\left(\boldsymbol{0},o^{*}\right)$ satisfying $R_{t}^{o^{*}}<1$ 
 are locally exponentially stable;
 \item All the dissensus-healthy equilibria $\left(\boldsymbol{0},o^{*}\right)$ satisfying $R_{t}^{o^{*}}>1$ 
 are unstable;
 \item The consensus-healthy equilibrium $\left(\boldsymbol{0},-0.5\boldsymbol{e}\right)$ is unstable.
\end{enumerate}
\end{theorem}


Theorem \ref{thm:case_3} implies that the local stability of the healthy equilibria 
depends on $R_{t}^{o^{*}}$. Moreover, one might never find a locally stable healthy equilibria, if no opinion state of the dissensus-equilibria satisfies Case 1) in Theorem \ref{thm:case_3}. \bshe{Theorem \ref{thm:case_3}  implies that, without interfering the opinions, the opinions of all communities cannot reach consensus at the healthy equilibrium, since the consensus-healthy equilibrium is unstable. Further, from Corollary \ref{cor:unique}, that the consensus-healthy state is the unique equilibrium for opinion consensus, we have that in the moderate virus case, the communities will not reach consensus in the absence of controlling the opinions.}


\bshe{Moreover, Theorem \ref{thm:case_3} states that the epidemic cannot be eradicated while all the communities are ignoring the epidemic. Further, if all the dissensus-healthy equilibria are unstable, i.e, $R_{t}^{o^{*}}>1$, $\forall (\boldsymbol{0},o^{*})$, the epidemic cannot reach a healthy state in the absence of control strategies. For the severe virus case, the highest level of awareness towards the seriousness of the epidemic cannot ensure the epidemic reaches a healthy state. However, with a moderate virus, to ensure the epidemic is eradicated, control strategies can be leveraged to reshape the opinions of the communities to maintain the opinion states above a certain threshold, such that the Opinion-Dependent Effective Reproduction Number is always below 1. Hence, the epidemic can be eradicated. }

To analyze the stability of the healthy equilibria under \bshe{interference on} opinion states, 
\bshe{we introduce threshold opinion vectors.
Since $R^{o}_t$ is related to $o(t)$, we can always find an opinion vector in $[-0.5, 0.5]^{n}$ such that $R^{o}_t\leq1$ or $R^{o}_t>1$, based on the definition of $R^{o}_t$,
and the moderate virus assumption that $R_{\max}>1$ and $R_{\min}<1$. The threshold opinion vectors are defined to analyze the system under the condition that all the opinion vectors are either smaller or greater than the threshold opinion vector. From the property of $R^{o}_t$ that $o(t_0)\leq o(t_1)$ leads to $R^{o}_{t_0} \geq R^{o}_{t_1}$ given by Proposition \ref{prop:Spe_r}, we have the following corollary.}

\begin{corollary}
\label{compare}
\bsheup{If $R_{\min}<1$ and $R_{\max}>1$, there must exist one threshold opinion vector $\bar{o}$ 
such that $R_{t}^{\bar{o}}=1$.}
%
\end{corollary}

\bsheup{\bs{Corollary} \ref{compare} is a direct result of Proposition \ref{prop:Spe_r}; 
note that there could be more than one $\bar{o}$ satisfying Corollary \ref{compare}.
Further, 
if 
$o(t)>\bar{o}$, $\forall t$, then $R_{t}^{o}<1$. Instead, if $o(t)<\bar{o}$, $\forall t$, then $R_{t}^{o}>1$. 
From Corollary \ref{compare}, we can capture the stability of the healthy equilibria in Theorem \ref{thm:case_3} through $\bar{o}$.
}
\begin{corollary}
\label{cor:threshold}
For the system in \eqref{eq:epi_op}, if $R_{\min}<1$ and $R_{\max}>1$, the dissensus-healthy equilibria  $(\boldsymbol{0},o^{*})$ satisfying $o^{*}\gg\bar{o}$ are locally exponentially stable, while the equilibria $(\boldsymbol{0},o^{*})$ satisfying  $o^{*}\ll\bar{o}$ are unstable\phil{.}
\end{corollary}

\begin{remark}
Case 3) of Theorem \ref{thm:case_3}  and Corollary \ref{cor:threshold} imply that, 
when the epidemic is 
moderate, the epidemic cannot be eradicated 
if
all communities ignore
it ($o^{*}=-0.5\boldsymbol{e}$) or
do not
treat
it  seriously enough ($o^{*}\ll\bar{o}$). Further,  Case 1) of Theorem \ref{thm:case_3} and Corollary \ref{cor:threshold} indicate that the epidemic will disappear when all communities 
believe that the epidemic is 
severe past a
certain degree
($o^{*}\gg\bar{o}$).
\end{remark}

\subsubsection{Stubborn Communities}

\bshe{After bridging the gap between the behavior of the healthy equilibria and the opinion threshold \bshe{vector} $\bar{o}$, we propose eradication strategies by leveraging the idea of reshaping the opinion formation of the communities through control strategies.} 
Note that the system in (\ref{eq:epi_op}) might have 
no stable healthy equilibrium. Assuming the system in (\ref{eq:epi_op}) 
has
at least one locally stable 
healthy
equilibrium,
from
Corollary \ref{compare}, there must exist an $\bar{o}$, s.t. $o^{*}\geq \bar{o}$. Therefore, to eradicate the epidemic, we can employ external influence on 
the communities to drive $o(t)$ above~$\bar{o}$. One method is to consider the existence 
of stubborn communities in (\ref{eq:op_c}), i.e., the opinions of the stubborn communities are not influenced by their neighbors \cite[Eq. (7)]{rahmani2009controllability}. 
Then, the following theorem captures the behavior of (\ref{eq:epi_op}) with stubborn communities. 



\begin{theorem}
\label{thm:stubborn_leader}
For 
the system in 
\eqref{eq:epi_op}, if $R_{\min}<1$ and $R_{\max}>1$, the 
stubborn communities driving 
$o(t)\gg\bar{o}$, $\forall t\geq 0$, will ensure 
\phil{that}
the system in \eqref{eq:epi_op} 
converges
to the set of healthy equilibria.
\end{theorem}

The proof of Theorem \ref{thm:stubborn_leader} is similar to the proof of Theorem~\ref{thm: stability_case1}, except that the opinion states are maintained 
above the threshold \bshe{vector} $\bar{o}$. 

\begin{remark}
Theorem \ref{thm:stubborn_leader} states, when the epidemic is moderate, we can eradicate 
it
by 
\bs{selecting} stubborn communities to drive the opinions of all the communities above the threshold vector $\bar{o}$. 
The situation implies that, by broadcasting the severity of the epidemic to some target communities, we can influence all the communities' beliefs toward the seriousness of the epidemic. In particular, when all communities' opinions are driven above a threshold vector, i.e., all the communities consider the epidemic somewhat serious, 
they will
take 
the
proper actions to end the epidemic. 
\end{remark}

Theorem \ref{thm:stubborn_leader} shows the role of stubborn communities in epidemic suppression. 
However, 
optimally
selecting the proper stubborn communities and designing external control signals to influence the stubborn communities 
is
challenging.
Hence, by Proposition \ref{prop:Spe_r}, we explore a particular method that considers stubborn communities with fixed opinion states equaling to $0.5$, $\forall t$. 
\phil{The following}
result 
provides
a way of selecting extreme 
stubborn communities for a particular case.

\begin{corollary}
\label{coro:method_2} Consider an opinion vector $\vec{o}$
with both positive and negative entries, $\vec{o}_{i}\in\left\{ -0.5,0.5\right\} $,
$\forall i\in[n]$. \bs{ If $\exists \vec{o}$, s.t. $R_{t}^{\vec{o}}<1$, the system in \eqref{eq:epi_op} can reach a healthy state by setting
$o_{i}(t)=0.5$, $\forall t\geq 0$, $\forall i$ satisfying $\vec{o}_{i}=0.5$. }
\end{corollary}
Corollary \ref{coro:method_2} reveals that, if we can find an $R_{t}^{\vec{o}}<1$, under the condition that the opinions of all communities are controlled at the extreme
beliefs, then $R_{t}^{o}$ will not exceed one after letting the communities with negative extreme opinions evolve freely, while maintaining the communities with positive extreme opinions the same.
Corollary \ref{coro:method_2} offers a way of selecting communities to make them stubborn 
in order
to suppress the epidemic. The elements of the opinion vector $\vec{o}_{i}\in\left\{ -0.5,0.5\right\}$ can be adjusted to generate different combinations of stubborn communities and opinions, e.g., exploring stubborn communities through Corollary \ref{coro:method_2} with the condition $\vec{o}_{i}\in\left\{ -0.5,o\right\}$, \bs{with $o=0.5\alpha$, $\alpha\in(-1,1]$ being the stubborn state.} 
\begin{remark}
Corollary \ref{coro:method_2} reveals the role of stubborn communities in determining the behavior of the epidemic. \bsheupp{As discussed in this section, when viruses are either mild or severe, without changing the opinions of all communities through control strategies, the epidemic caused by mild viruses will disappear even with communities ignoring the virus, while the epidemic caused by severe viruses will not be eradicated even when 
all communities take various actions caused by the highest awareness toward the epidemic. 
In the moderate virus case, where 
extreme 
opinions from both sides (ignoring the epidemics or treating the epidemic extremely seriously) play important roles in determining the spread of the epidemic, positive stubborn opinions will drive the overall opinion states higher, leading to lower $R^{o}_{t}$. 
On the other hand, stubborn communities with negative opinions will increase the $R^{o}_{t}$, making it harder for the communities to eradicate epidemics. Additionally, the existence of the epidemic may lead to dissensus on the seriousness of the epidemic, which is modeled by the signed switching opinion network structure.}
\end{remark}
\section{Simulations}
\begin{figure}
\begin{centering}
\includegraphics[scale=0.15]{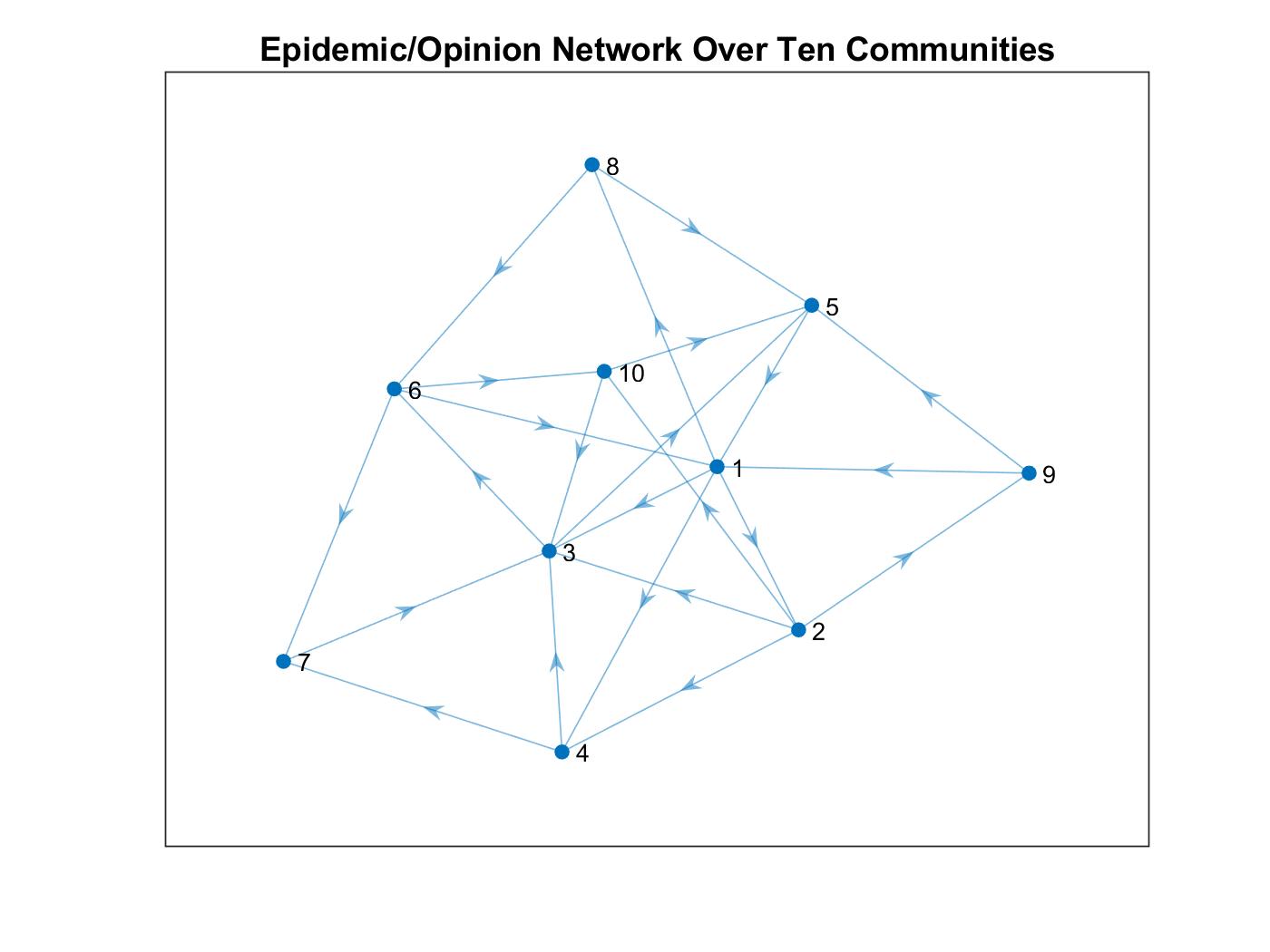}
\par\end{centering}
\centering{}\caption{The graph for the epidemic and opinion interactions.
}
\label{fig_graph}
\end{figure}

In this section, we illustrate the main results 
through the following examples. Consider an epidemic process spreading over ten communities, with the epidemic and opinion spreading through the same network satisfying Assumption \ref{A1}, captured by the graph $G$ in Fig. \ref{fig_graph}. Note that we use the same 
graph structure in $G$ to capture the epidemic and opinion graphs to simplify the simulation, and our results still apply to communities with different epidemic and opinion interactions. \bshe{Note that the goal of the simulations is to illustrate the theoretical results qualitatively (e.g., whether the epidemic disappears eventually or not or the opinion reaches consensus or dissensus) instead of quantitatively (e.g., the endemic states of each community/the exact opinion formation under dissensus).}

\begin{figure}
\begin{centering}
\includegraphics[scale=0.65]{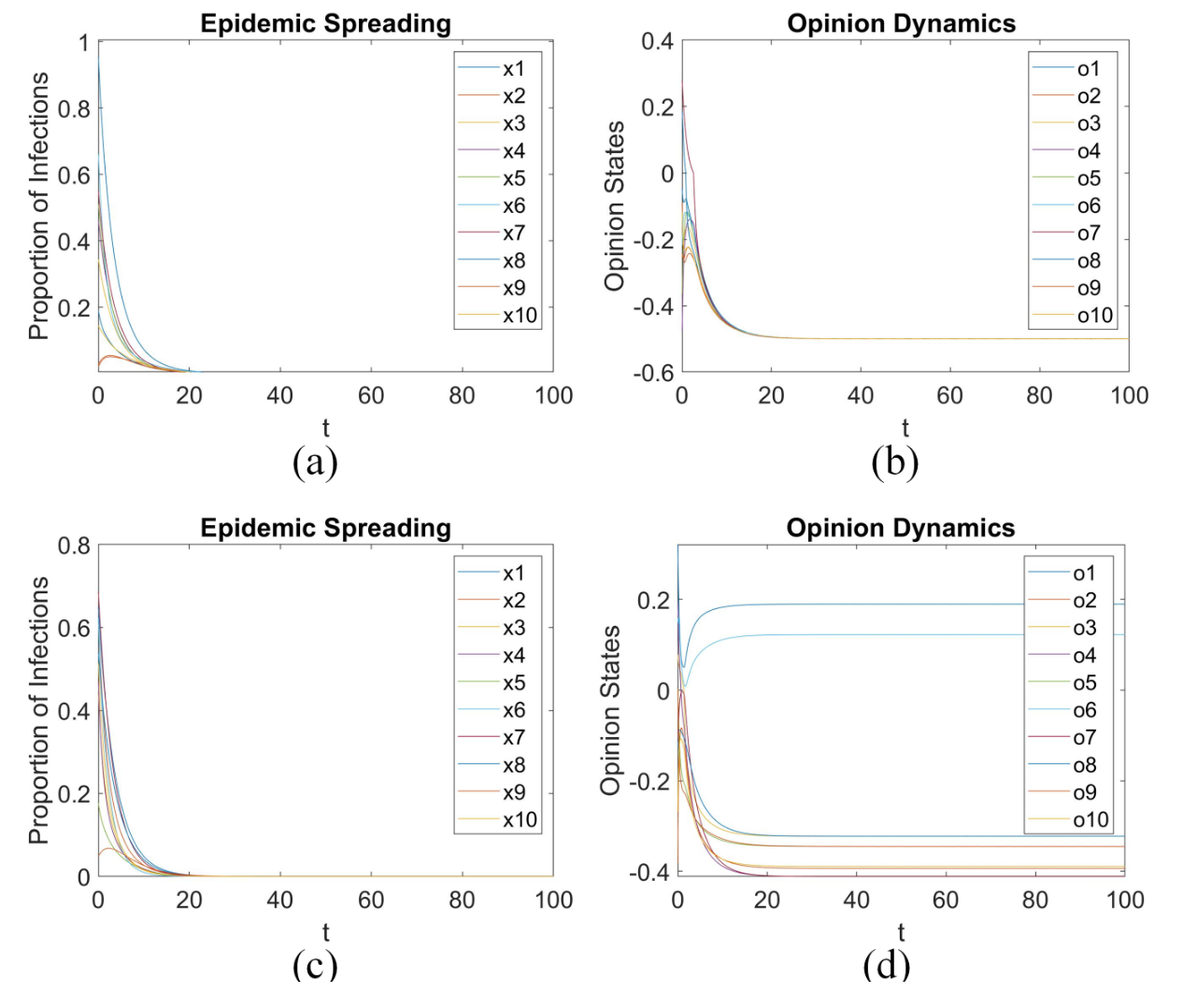}
\par\end{centering}
\centering{}\caption{ Under the condition $R_{\min}=0.174$, $R_{\max}=0.381$, with the graph of ten communities from Fig. \ref{fig_graph}, 
the communities reach the healthy state: (a) epidemic states converge to zero, (b) opinion states reach consensus, (c) epidemic states converge to zero, (d) opinion states reach dissensus.} 

\label{fig_case1}
\end{figure}

First, we consider the case that the epidemic is mild, which indicates 
$R_{\max}\leq 1$. By generating parameters of the infection and healing rates randomly, we obtain $R_{\min}=0.174$, $R_{\max}=0.381$. 
Consistent with
Proposition \ref{prop:case1} and Theorem \ref{thm: stability_case1}, when $R_{\max}\leq 1$, i.e., the epidemic is mild, all the communities reach healthy states with zero infections, 
illustrated in
Fig. \ref{fig_case1} (a) and (c). Further, \bs{under the condition that the epidemic will eventually disappear}, Theorem \ref{thm: stability_case1} states that opinions of the communities either reach the consensus-healthy state $\left(\boldsymbol{0},-0.5\boldsymbol{e}\right)$, where all the communities agree that the epidemic is not serious, 
illustrated in
Fig. \ref{fig_case1} (a) and (b), or a dissensus-healthy state $\left(\boldsymbol{0},o^{*}\right)$, with the communities holding both positive and negative opinion states toward the epidemic, 
illustrated in
Fig. \ref{fig_case1} (c) and (d). \bshe{
Additionally, it is implied from the simulation that opinion consensus or dissensus after the epidemic disappears is dependent on the initial infected proportion and initial awareness towards the seriousness of each community. 
By comparing Fig. \ref{fig_case1} (b) and (d), if the opinion of each community drops below $0$ early, all the communities will reach consensus on the seriousness of the epidemic. The conditions for reaching consensus and dissensus are worth exploring in future work.}

\begin{figure}
\begin{centering}
\includegraphics[scale=0.30]{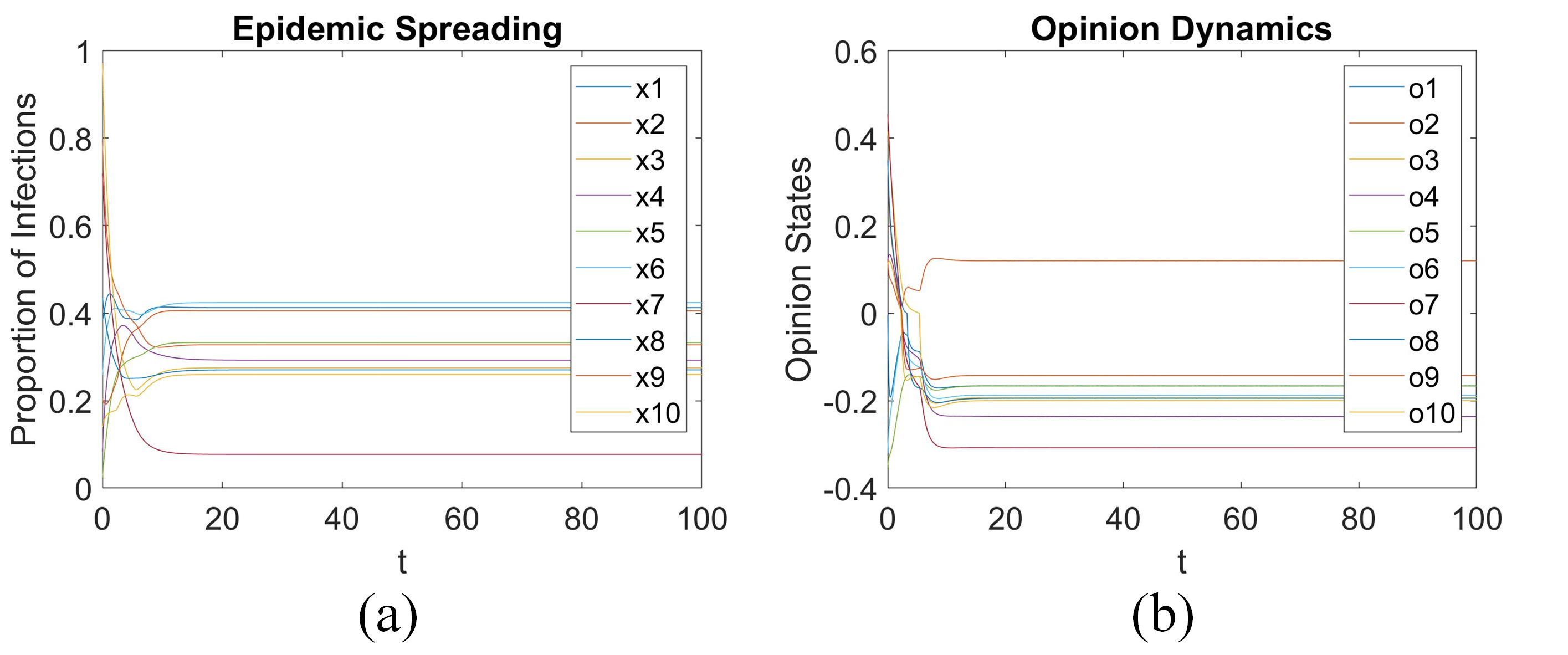}
\par\end{centering}
\centering{}\caption{ 
The communities under the condition $R_{\min}=1.63$, $R_{\max}=2.84$, with the graph of ten communities from Fig. \ref{fig_graph}, reach 
a dissensus-endemic state: (a) epidemic states reach endemic equilibrium,
(b) opinion states reach dissensus.
}

\label{fig_case2}
\end{figure}

Fig. \ref{fig_case2} illustrates the situation 
where
the epidemic is severe, 
characterized by $R_{\min}>1$. Through randomly generating parameters satisfying the condition, we have $R_{\min}=1.63$, $R_{\max}=2.84$. 
Consistent with
Lemma \ref{lem:unstable}, 
none
of the communities can 
ever reach
a
healthy state. Fig. \ref{fig_case2} (a) also implies the existence of 
an
endemic equilibrium, 
illustrating
Theorem \ref{thm:case2}. 
Fig. \ref{fig_case2} (b) 
shows that all 
the ten communities reach dissensus, \bshe{since as shown in Corollary \ref{cor:unique}, the unique case that all communities reach consensus on the seriousness of the epidemic is dependent on the condition that the epidemic dies out. Therefore, when the epidemic reaches an element in the set of endemic states, as shown in Fig. \ref{fig_case2} (a), the opinion must reach dissensus as in Fig. \ref{fig_case2} (b). Further, when the epidemic becomes endemic, the opinion of each community at the equilibria can be obtained by solving (\ref{eq:epi}) and (\ref{eq:opi})}. Note that the continuity at the opinion-switching points and the Lipschitz continuity between the opinion-switching points can be observed from the example.

\begin{figure}
\begin{centering}
\includegraphics[scale=0.31]{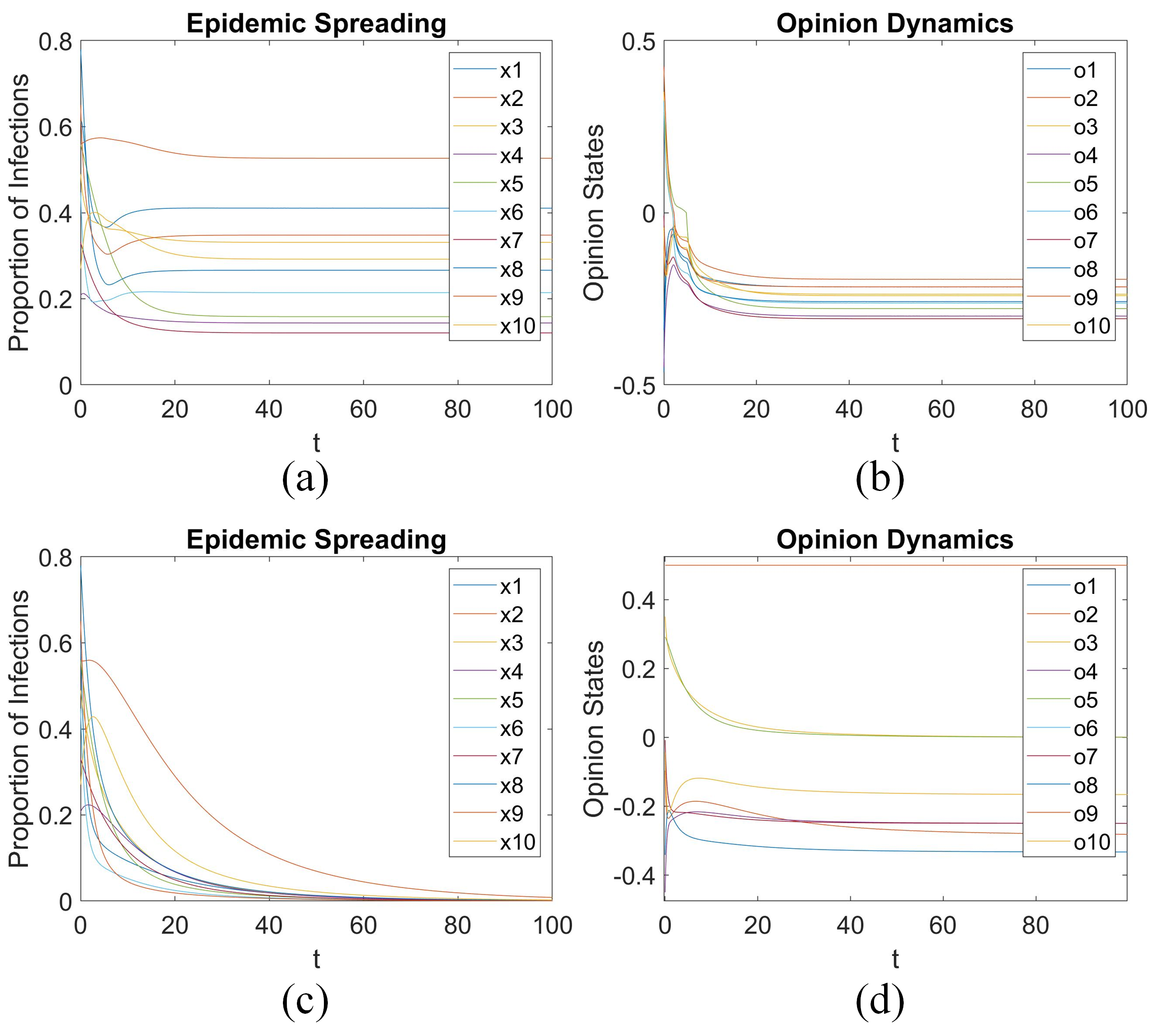}
\par\end{centering}
\centering{}\caption{ All the communities under the condition $R_{\min}=0.51$, $R_{\max}=2.61$, with the graph of ten communities from Fig. \ref{fig_graph}: (a) epidemic states reach endemic equilibrium,
(b) opinion states without stubborn opinions, 
(c) epidemic states converge to zero,
(d) opinion states with stubborn opinions.
}

\label{fig_case3}
\end{figure}

Lastly, we consider the case 
where
the epidemic is moderate. 
Thus, we generate parameters leading to $R_{\min}=0.51<1$, $R_{\max}=2.61>1$ to characterize this situation. 
As described in Theorem \ref{thm:case_3}, the stability of the healthy equilibria depends on the opinion states. More importantly, from Theorem \ref{thm:stubborn_leader} and Corollary \ref{coro:method_2}, we can 
appropriately
select 
stubborn communities to eradicate the epidemic. 
Fig. \ref{fig_case3} illustrates the role of stubborn communities in epidemic suppression,
showing
the same system 
under
two different settings. The system captured by Fig. \ref{fig_case3} (a) and (b) reaches a dissensus-endemic state. With the exact same conditions, we fix the opinion states of the communities 1, 6, and 9 as $\left[o\left(t\right)\right]_{1}=\left[o\left(t\right)\right]_{6}=\left[o\left(t\right)\right]_{9}=0.5$, $\forall t\geq 0$, which means communities 1, 6, and 9 always believe that the epidemic is extremely severe, and the opinions of other communities will not impact 
their beliefs in the severity of the epidemic. Meanwhile, communities 1, 6, and 9 keep broadcasting the information that the epidemic is very severe to their neighbors. Through this setting, compared to the same system captured by Fig. \ref{fig_case3} (a) and (b), Fig. \ref{fig_case3} (c) and (d) show that all the communities reach the healthy state, illustrating the results derived in Theorem \ref{thm:case_3}, Theorem \ref{thm:stubborn_leader}, and Corollary \ref{coro:method_2}, that appropriate selection of stubborn communities, broadcasting their cautious opinions to other communities, can suppress the epidemic. 

\section{Conclusion}
This work studies the mutual influence between epidemic spreading and opinion dissemination over 
connected communities. By defining an Opinion-Dependent 
Reproduction Number, our work reveals the behavior
of the Epidemic-Opinion model in (\ref{eq:epi_op}).
Our work also illustrates the role of stubborn communities in epidemic eradication. The results of this work pave the way for a more detailed analysis of the extended models. In this work, the stability analysis of the endemic states  under the condition that $R_{\min}>1$ still needs to be further explored. The study of the endemic states could reveal the impact of 
opinions on the most serious epidemics.  
Next, except for using stubborn opinions, one could consider control design to shape the opinions of the communities 
to eradicate the epidemic.
Further extensions of the results in this work
consist of validating
the opinion-dependent sign switching network model with real-world data
and extending the ideas to couple opinion dynamics with
the SIR (susceptible-infected-recovered) model. 
\bshe{
Additionally, diseases including SARS-CoV-2 have an incubation period that cannot be ignored \cite{prem2020,kucharski2020}, which is not considered in our current model. The incubation period, usually captured by the exposed compartment in a compartmental model, will bring new challenges in analyzing opinion formation on epidemic spreading, since the infected population is not aware of its infection during the incubation period. Future work on studying SEIR (susceptible-exposed-infected-recovered) epidemic models coupled with opinion dynamics would therefore be of value.
}
\bibliographystyle{IEEEtran}
\typeout{} 

\bibliography{IEEEabrv,bib}

%

\begin{figure}[h]
\begin{centering}
\includegraphics[scale=0.31]{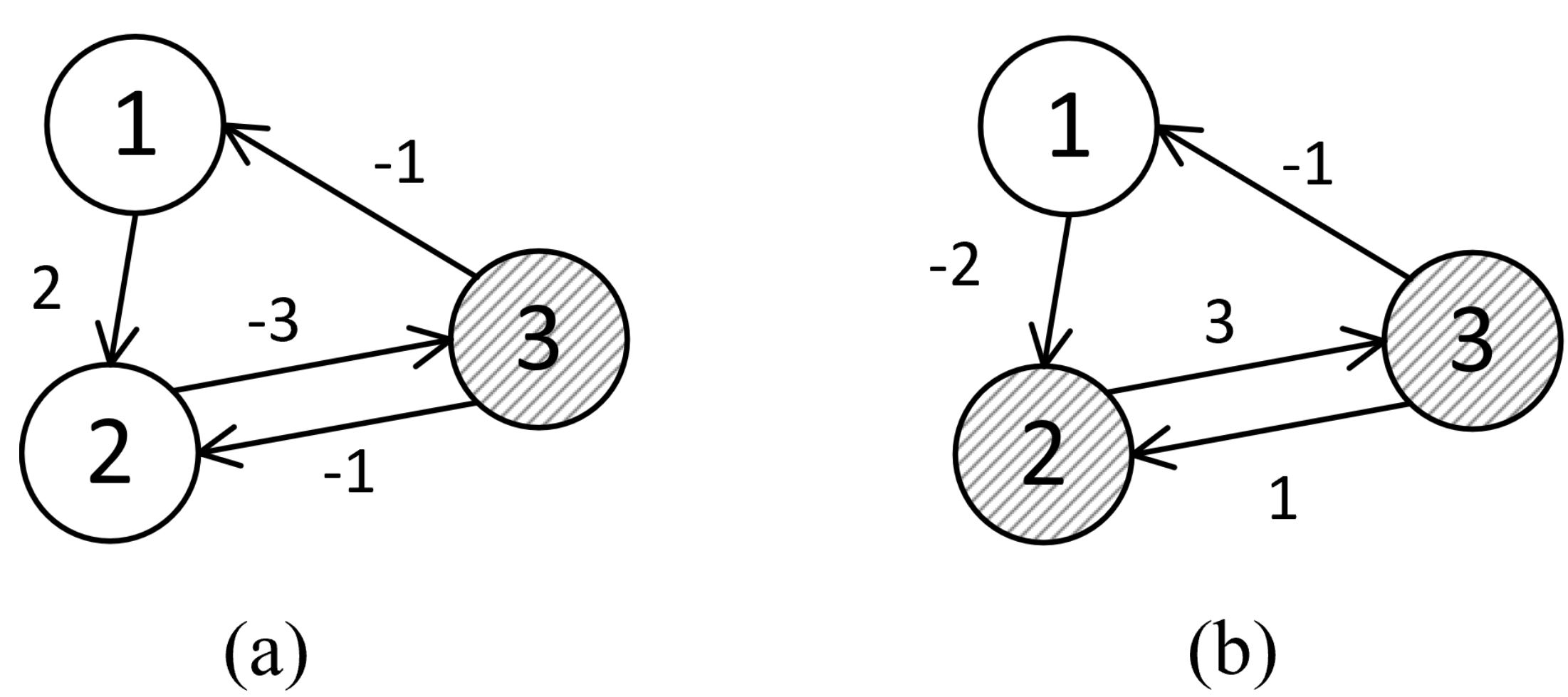}
\par\end{centering}
\centering{}\caption{ 
\bsheup{Opinion spreading networks: (a) Opinion formation and the corresponding network structure at $t_0$,
(b) Opinion formation and the corresponding network structure at $t_1$.}}
\label{fig_example}
\end{figure}

\appendix
\vspace{2ex}
\bsheup{\textit{Example of Eq. \eqref{eq:opi_sb}}: Consider opinion interactions captured by the graph with three nodes in Fig. \ref{fig_example}. The positive and negative opinion states are represented by unshaded and shaded nodes, respectively. At time $t_0$, the unsigned Laplacian matrix $L_u$ of the graph in Fig. \ref{fig_example} (a) and the time-varying gauge transformation matrix $\Phi(o(t_{0}))$ defined in Lemma \ref{Lem:SB} are given by
\begin{align*}
\bar{L}_{u}=\left[\begin{array}{ccc}
1 & 0 & -1\\
-2 & 3 & -1\\
0 & -3 & 3
\end{array}\right], 
\Phi(o(t_{0}))=\left[\begin{array}{ccc}
1 & 0 & 0\\
0 & 1 & 0\\
0 & 0 & -1
\end{array}\right],
\end{align*}
}
\\
\bsheup{
respectively. Since node 1 and node 2 share their opinions cooperatively, while node 3 shares its opinion with node 1 and node 2 antagonistically, we have $[\Phi(o(t_{0}))]_{11}=[\Phi(o(t_{0}))]_{22}$, $\Phi(o(t_{0}))]_{33}=-1$. Based on Lemma \ref{Lem:SB}, the signed Laplacian matrix of opinion formation in Fig. \ref{fig_example} (a) is $\bar{L}(o(t_0))=\Phi(o(t_{0}))\bar{L}_{u}\Phi(o(t_{0}))$. At time $t_1$, the opinion of node 2 drops below zero, which enables node 2 to share a positive edge with node 3, and a negative edge with node 1. Thus, the signed Laplacian matrix of the graph in Fig. \ref{fig_example} (b) is given by $\bar{L}(o(t_1))=\Phi(o(t_{1}))\bar{L}_{u}\Phi(o(t_{1}))$, where the
entries of the gauge transformation matrix at $t_1$ are given by $[\Phi(o(t_{1}))]_{11}=1, [\Phi(o(t_{1}))]_{22}=[\Phi(o(t_{1}))]_{33}=-1$.}

\vspace{2ex}
\textit{Proof of Lemma \ref{lem:Inv}:}
Consider the system 
captured by \eqref{eq:epi_c} and \eqref{eq:op_c}. Note that the system in \eqref{eq:epi_c} is a group of
polynomial ODEs over the compact set $\left[0,1\right]^{n}$. Between the switching points, each subsystem in \eqref{eq:op_c} is a group of polynomial ODEs over the compact set $\left[-0.5,0.5\right]^{n}$. Therefore, for each subsystem of \eqref{eq:op_c} paired with \eqref{eq:epi_c}, the system \eqref{eq:epi_c}
is Lipschitz on $\left[0,1\right]^{n}$ and each subsystem of \eqref{eq:op_c} is Lipschitz on $\left[-0.5,0.5\right]^{n}$. It can be verified that the solutions at the switching
points of \eqref{eq:op_c} are continuous. Hence,
the solutions $x_{i}\left(t\right)$ and $o_{i}\left(t\right)$ of \eqref{eq:epi_c} and \eqref{eq:op_c} are continuous, $\forall i\in\left[n\right]$, respectively.

Suppose there is an index $i\in\left[n\right]$ such that $x_{i}\left(t\right)$
is the first state to reach zero at $t_{0}$, while the rest of the
states $x_{j}(t_0)\in \mathrm{Int}\left[ 0,1\right]^{n}$ and $o_{j}(t_0)\in \mathrm{Int}\left[-0.5,0.5\right]^{n}$,
$\forall j\in\left[n\right]$, $i\neq j$. Based on \eqref{eq:epi} and Assumption $\ref{A1}$,
\[
\dot{x}_{i}\left(t_{0}\right)=\sum_{j\in \mathcal{N}_{i}}\left[\beta_{ij}-\left(\beta_{ij}-\beta_{\min}\right)\left(o_{i}(t_0)+0.5\right)\right]x_{j}\left(t_{0}\right)\ge0.
\]
Hence, $\dot{x}_{i}\left(t_{0}\right)\ge0$ indicates ${x}_{i}\left(t_{0}\right)$
cannot drop below zero when being the first to reach zero. 
The same statements hold for the situations 
where
more than one of the epidemic
states reach 
$\partial\left[0,1\right]^{n}$, simultaneously. Following the same procedure, we can verify that $x\left(t\right)\leq\boldsymbol{e}$, $\forall t\geq 0$.
Consider the opinion dynamics in \eqref{eq:op_c}. 
From the same analysis, it can be verified
that $o_{i}(t)\in\left[-0.5,0.5\right]$, $\forall t\geq 0$ , $\forall i\in \left[n\right]$.
\qed

\vspace{2ex}
\textit{Proof of Theorem \ref{thm:equi_basic}:}
\baike{We} first show that the healthy state $x^{*}=\boldsymbol{0}$ can only pair with the unique consensus state $o^{*}=-0.5\boldsymbol{e}$.
\baike{Recall} the definition of $\Phi\left(o\left(t\right)\right)$ \baike{in Section \ref{sec:model_o};} if $o_{i}^{*}=o_{j}^{*}$, $\forall i,j\in\left[n\right]$, then $\Phi\left(o^{*}\right)=\pm I$. 
\baike{From} \eqref{eq:op_c},
the equilibria of the opinion dynamics 
satisfy
\begin{equation}
    -\left(\bar{L}_{u}+I\right)o^{*}=0.5\boldsymbol{e}.
    \label{eq:op_equi}
\end{equation}
Thus we have that $-0.5\boldsymbol{e}$ is an eigenvector of the matrix
$-\left(\bar{L}_{u}+I\right)$ paired with the largest eigenvalue $-1$. \baike{From Remark \ref{re:basics}}, $-\left(\bar{L}_{u}+I\right)$ is nonsingular, 
and thus
$o^{*}=-0.5\boldsymbol{e}$ is the 
\baike{unique}
solution of \baike{\eqref{eq:op_equi}\phil{.}}
Therefore, $\left(x^{*}=\boldsymbol{0}, o^{*}=-0.5\boldsymbol{e}\right)$ is the \baike{unique} consensus-healthy equilibrium of \eqref{eq:epi_op}.

For dissensus-healthy states, $\left(x^{*}=\boldsymbol{0}, o^{*}\right)$, if $o^{*}\gg\boldsymbol{0}$ or $o^{*}\ll\boldsymbol{0}$, \baike{which implies $\Phi\left(o^{*}\right)=\pm I$}, the equilibrium of the opinion dynamics in \eqref{eq:op_c} becomes $-\left(\bar{L}_{u}+I\nonumber\right)o^{*}=0.5\boldsymbol{e}
$, which has only $-0.5\boldsymbol{e}$, the consensus state, as its solution. Therefore, $o^{*}$ in $\left(x^{*}=\boldsymbol{0}, o^{*}\right)$ must have both positive and negative entries. 
\bs{Based on 
the fact that $\left(\Phi\left(o\left(t\right)\right) {\bar{L}}_{u}\Phi\left(o\left(t\right)\right)+I\right)$ is a nonsingular matrix,} the equation
%
\begin{equation}
\left(\Phi\left(o^{*}\right)\bar{L}_{u}\Phi\left(o^{*}\right)+I\right)o^{*}=x^{*}-0.5\boldsymbol{e},
\label{eq:equi_opinion}
\end{equation}
has a unique solution for each $\left(\Phi\left(o^{*}\right)\right)$, given by $o^{*}=\left(\Phi\left(o^{*}\right) \bar{L}_{u}\Phi\left(o^{*}\right)+I\right)^{-1}\left(-0.5\boldsymbol{e}\right)$, \bs{when $x^{*}=\boldsymbol{0}$}.
Now we show that each solution must satisfy $0.5\boldsymbol{e}\geq o^{*}\geq\boldsymbol-0.5\boldsymbol{e}$.

Assume that $\left[\Phi\left(o^{*}\right)\right]_{ii}=-1$, $\forall i\in\{1,\ldots,m\}$ and $\left[\Phi\left(o^{*}\right)\right]_{jj}=1$,  $\forall j\in\{m+1,\ldots,n\}$.
Let $\bar{L}=\Phi\left(o^{*}\right)\bar{L}_{u}\Phi\left(o^{*}\right)$. Without loss of generality, suppose to the contrary that $o_{1}^{*}<-0.5$. Based on the assumption,
$o_{i}^{*}<0$, $\forall i\in\{2,\ldots,m\},$ while $o_{j}^{*}\geq0$, $\forall j\in \{m+1,\ldots,n\}$. Considering the first row of \eqref{eq:equi_opinion}, we have
\[
-\left|\left[\bar{L}\right]_{11}\right|o_{1}^{*}-o_{1}^{*}+\sum_{i=2}^{m}\left|\left[\bar{L}\right]_{1i}\right|o_{i}^{*}-\sum_{j=m+1}^{n}\left|\left[\bar{L}\right]_{1j}\right|o_{j}^{*}=0.5.
\]
For $\left|\left[\bar{L}\right]_{11}\right|=\sum_{k=2}^{n}\left|\left[\bar{L}\right]_{1k}\right|$, we have
\begin{align*}
-\sum_{i=2}^{m}\left|\left[\bar{L}\right]_{1i}\right|o_{1}^{*}-\sum_{j=m+1}^{n}\left|\left[\bar{L}_{1j}\right]\right|o_{1}^{*}\\
+\sum_{i=2}^{m}\left|\left[\bar{L}\right]_{1i}\right|o_{i}^{*}-\sum_{j=m+1}^{n}\left|\left[\bar{L}_{1j}\right]\right|o_{j}^{*}-o_{1}^{*} & =0.5.
\end{align*}
Note that if $o_{i}^{*}\geq o_{1}^{*}$, $\forall i\in\{2,\ldots,m\}$, and $o_{j}^{*}\leq -o_{1}^{*}$, $\forall j\in\{m+1,\ldots,n\}$, the left side of the equation above must be greater than $0.5$. Therefore, there must exist at least one $o_{i}^{*}$, $i\in\{2,\ldots,m\}$, satisfying $o_{i}^{*}<o_{1}^{*}$
 and/or at least one $o_{j}^{*}$, $j\in\{m+1,\ldots,n\}$, satisfying $o_{j}^{*}>-o_{1}^{*}$.
Suppose that $o_{2}^{*}<o_{1}^{*}$, then for the second row of \eqref{eq:equi_opinion}, the same statement holds, that there must exist at least one $o_{i}^{*}$, $i\in\{3,\ldots,m\}$, satisfying $o_{i}^{*}<o_{2}^{*}<o_{1}^{*}$, and/or at least one $o_{j}^{*}$, $j\in\{m+1,\ldots,n\}$, satisfying $o_{j}^{*}>-o_{2}^{*}>-o_{1}^{*}$. Following this procedure, for the last row corresponding to the last entry in $o^{*}$, we can no longer find any entries satisfying the condition. Therefore, no solution of \eqref{eq:equi_opinion} can be smaller than $-0.5$. A similar process can be applied to show that no solution of the equation \eqref{eq:equi_opinion} can have an element larger than~$0.5$.  

Therefore, for each sign pattern of $\Phi\left(o^{*}\right)$, the system in \eqref{eq:epi_op} has one unique dissensus-healthy state $o^{*}=\left(\Phi\left(o^{*}\right) \bar{L}_{u}\Phi\left(o^{*}\right)+I\right)^{-1}\left(-0.5\boldsymbol{e}\right)$.
\qed

\vspace{2ex}
\textit{Proof of Corollary \ref{cor:unique}:}
Theorem \ref{thm:equi_basic} shows that the consensus-healthy state $\left(x^{*}=\boldsymbol{0}, o^{*}=-0.5\boldsymbol{e}\right)$, is \baike{the} unique 
equilibrium when $x^{*}=\boldsymbol{0}$. Therefore, we need to show \baike{that} there exists no consensus-endemic state 
that is
an equilibrium. 
\baike{Suppose to the contrary that there exists an equilibrium} 
$o$, s.t. $o=\alpha\boldsymbol{e}$, $\alpha\in\left(-0.5,0.5\right]$. Based on \eqref{eq:equi_opinion}, $x=\alpha\boldsymbol{e}+0.5\boldsymbol{e}$. Substituting 
$\left(x,o\right)$ in \eqref{eq:op_c}, we have

\begin{align*}
\dot{o} & =\alpha\boldsymbol{e}+0.5\boldsymbol{e}-\left(I\bar{L}_{u}I+I\right)\times0.5\boldsymbol{e}-0.5e\\
 & =\alpha\boldsymbol{e}-0.5\boldsymbol{e}\leq\boldsymbol{0}.
\end{align*}
The inequality becomes an equality only under the condition that $\left(x=\boldsymbol{e}, o=0.5\boldsymbol{e}\right)$. However, by substituting $\left(x=\boldsymbol{e}, o=0.5\boldsymbol{e}\right)$ into \eqref{eq:epi_c}, we have $\dot{x} <\boldsymbol{0}$.
Thus, $\left(x,o\right)$ cannot be an equilibrium of \eqref{eq:epi_op}. Therefore, by contradiction, the healthy-consensus state $\left(x^{*}=\boldsymbol{0}, o^{*}=-0.5\boldsymbol{e}\right)$ is the unique equilibrium with consensus in opinions.
\qed

\vspace{2ex}
\textit{Proof of Lemma \ref{lem:equi_non}:}
\baike{
First we show that $x^{*}\gg\boldsymbol{0}$. Suppose to the contrary that $\exists i\in\left[n\right]$, s.t. $x^{*}_{i}=0$, while $x^{*}_{j}\neq 0$, for all $j\in\left[n\right]$, $\forall i\neq j$. Based on the proof of Lemma \ref{lem:Inv}, if $\dot{x}^{*}_{i}=0$ at $x^{*}_{i}=0$, we have $x^{*}_{j}=0$, for all other $j\in\left[n\right]$.
The  same  statements hold for the situations where more than one of the elements in $x^{*}$ equal to zero.
Therefore, $x^{*}$ must satisfy $x^{*}\gg\boldsymbol{0}$.
We can apply the similar proof to show that $x^{*}\ll\boldsymbol{e}$, $-0.5\boldsymbol{e}\ll o^{*}\ll0.5\boldsymbol{e}$.}
\qed

\vspace{2ex}
\textit{Proof of Proposition \ref{prop:Spe_r}:} 1) Based on Assumption \ref{A1},
$D\left(o\left(t\right)\right)$ is
a \bs{positive definite} diagonal matrix 
and $B\left(o\left(t\right)\right)$
is an irreducible nonnegative matrix, $\forall t$. Hence, $D\left(o\left(t\right)\right)^{-1}B\left(o\left(t\right)\right)$
is an irreducible nonnegative matrix. Without loss of generality, consider the case where, $\exists o_{i}\left(t_{0}\right)<o_{i}\left(t_{1}\right)$, $i\in\left[n\right]$, $t_{1}>t_{0}>0$,
while 
$o_{j}\left(t_{0}\right)=o_{j}\left(t_{1}\right)$,
$\forall i,j\in\left[n\right]$, $i\neq j$. Recall that $D\left(o\left(t\right)\right)=D_{\min}+\left(D-\delta_{\min}I\right)\left(O\left(t\right)+0.5I\right)$ and
$B\left(o\left(t\right)\right)=B-\left(O\left(t\right)+0.5I\right)\left(B-B_{\min}\right)$.
Based on $o_{i}\left(t_{0}\right)<o_{i}\left(t_{1}\right)$, we have
$D_{ii}\left(o\left(t_{0}\right)\right)<D_{ii}\left(o\left(t_{1}\right)\right)$,
leading to $D_{ii}^{-1}\left(o\left(t_{0}\right)\right)>D_{ii}^{-1}\left(o\left(t_{1}\right)\right)$,
and $B_{i,:}\left(o\left(t_{0}\right)\right)>B_{i,:}\left(o\left(t_{1}\right)\right)$,
while the rest 
of $D^{-1}\left(o\left(t_{1}\right)\right)$
and $B\left(o\left(t_{1}\right)\right)$ 
are equal to $D^{-1}\left(o\left(t_{0}\right)\right)$
and $B\left(o\left(t_{1}\right)\right)$, respectively. Hence,
$o_{i}\left(t_{0}\right)<o_{i}\left(t_{1}\right)$ leads to
\[
\left[D\left(o\left(t_{0}\right)\right)^{-1}B\left(o\left(t_{0}\right)\right)\right]_{i,:}>\left[D\left(o\left(t_{1}\right)\right)^{-1}B\left(o\left(t_{1}\right)\right)\right]_{i,:}.
\]
From \baike{\cite[Thm. 2.7, and Lemma 2.4]{varga2009matrix_book},}
we have 
\[
\rho\left[D\left(o\left(t_{0}\right)\right)^{-1}B\left(o\left(t_{0}\right)\right)\right]>\rho\left[D\left(o\left(t_{1}\right)\right)^{-1}B\left(o\left(t_{1}\right)\right)\right],
\]
which means $R_{t_{0}}^{o}> R_{t_{1}}^{o}$. The proof holds for the situations where more than one opinion states in $o(t_0)$ are smaller than $o(t_1)$.
The same method can verify the case that $o\left(t_{0}\right)\geq o\left(t_{1}\right)$, then $R_{t_{0}}^{o}\leq R_{t_{1}}^{o}$.

2) This statement is 
two special cases of \bs{1)}. 
Since $-0.5\boldsymbol{e}\leq o\left(t\right)\leq0.5\boldsymbol{e}$, when $o\left(t\right)=o_{\min}=-0.5\boldsymbol{e}$, based on the first statement of Proposition \ref{prop:Spe_r},

\begin{align*}
\rho\left(D\left(o\left(t\right)\right)^{-1}B\left(o\left(t\right)\right)\right) & \leq\rho\left[D\left(o_{\min}\right)^{-1}B\left(o_{\min}\right)\right]\\
 & =\rho\left(D_{\min}^{-1}B\right)=R_{\max}.
\end{align*}
When $o\left(t\right)=o_{\max}=0.5\boldsymbol{e}$, 

\begin{align*}
\rho\left(D\left(o\left(t\right)\right)^{-1}B\left(o\left(t\right)\right)\right) & \geq\rho\left(D\left(o_{\max}\right)^{-1}B\left(o_{\max}\right)\right)\\
 & =\rho\left(D^{-1}B_{\min}\right)=R_{\min}.
\end{align*}

\qed

\vspace{2ex}
\textit{Proof of Proposition \ref{prop:case1}:}
Suppose to the contrary that there is an endemic state  $\left(x,o\right)$ as the equilibrium
of \eqref{eq:epi_op} under the condition that 
\baike{$R_{\max}\leq 1$}.
By Lemma \ref{lem:equi_non}, it must be true that $\boldsymbol{e}\gg x\gg\boldsymbol{0}$.
Since $\left(x,o\right)$ is an equilibrium, from \eqref{eq:epi_c},

\begin{align*}
\left(-D_{\min}+B\right)x & =XBx+\left(O+0.5I\right)\left(D-D_{\min}\right)x\\
 & \text{}+\left(I-X\right)\left(\left(O+0.5I\right)\left(B-B_{\min}\right)\right)x.
\end{align*}
\bs{By} Assumption 1, both 
$\left(B-B_{\min}\right)$ and $\left(I-X\right)\left(\left(O+0.5I\right)\left(B-B_{\min}\right)\right)$
are nonnegative and irreducible, and 
$\left(O+0.5I\right)\left(D-D_{\min}\right)$
is a \bs{positive definite} diagonal matrix. 
Hence, \bs{since} $x\gg\boldsymbol{0}$, we have
$XBx\gg\boldsymbol{0}$, $\left(O+0.5I\right)\left(D-D_{\min}\right)x\gg\boldsymbol{0}$, $\left(I-X\right)\left(\left(O+0.5I\right)\left(B-B_{\min}\right)\right)x\gg\boldsymbol{0}$.
Therefore, $\left(-D_{\min}+B\right)x\gg\boldsymbol{0}$. 

Recall that $\left(-D_{\min}+B\right)$ is an irreducible nonnegative matrix; from
\cite[Sec. 2.1 and Lemma 2.3]{varga2009matrix_book}, $s(-D_{\min}+B)>0$. However, \bs{by} \cite[Prop. 1]{bivirus}, $s(-D_{\min}+B)>0$ leads to $\rho(D_{\min}^{-1}B)=R_{\max}>1$, which contradicts the assumption of the proposition that $R_{\max}\le1$. Therefore, an endemic state $\text{\ensuremath{\left(x,o\right)}}$
cannot be an equilibrium of \eqref{eq:epi_op}
if $R_{\max}\le1$. 
\qed

\vspace{2ex}
\textit{Proof of Proposition \ref{prop:unstable}:}
We derive the Jacobian matrix $df_{x,o}$ of \eqref{eq:epi_op} evaluated at $\left(x,o\right)$ as follows:
%
\[
\left[\begin{array}{cc}
W(o)-\tilde{V}\left(x,o\right) & \text{\ensuremath{\text{\ensuremath{}}-\left(D-D_{\min}\right)\bs{X}-\left(I-X\right)\tilde{B}}}\\
I & -\left(\Phi\left(o\right)\bar{L}_{u}\Phi\left(o\right)+I\right)-\tilde{\Delta}
\end{array}\right],
\]
where $\tilde{V}\left(x,o\right)$ and $\tilde{B}$ are diagonal matrices
with 
the $i$th diagonal entries being the $i$th entries of the vectors
$\left(B-\left(O+0.5I\right)\left(B-B_{\min}\right)\right)x$
and $\left(B-B_{\min}\right)x^{2}$, respectively, and $\tilde{\Delta}=\left(\Delta\left(\bar{L}_{u}+I\right)\Phi\left(o\right){o}+\Phi\left(o\right) \left(\bar{L}_{u}+I\right)\Delta{o}\right)$, 
\phil{with the Dirac delta function $\theta\left(\cdot \right)$ and  $\Delta=\diag\left\{ 2\theta\left(o_{1}\right),\ldots,2\theta\left(o_{n}\right)\right\}$.}
From Corollary \ref{cor:not_0}, we have 
$o_{i}^{*}\neq0$, $\forall i\in\left[n\right]$, for all equilibria of \eqref{eq:epi_op}. Therefore, $\tilde{\Delta}=\boldsymbol{0}$ when evaluated at all the equilibria, due to $\theta\left(o_{i}\right)=0$, when $o_{i}\neq 0$, $\forall i\in\left[n\right]$.
We evaluate the Jacobian matrix at all healthy equilibria, $\left(x=\boldsymbol{0}, o^{*}\right)$,
\begin{equation}
df_{\boldsymbol{0}, o^{*}}=\left[\begin{array}{cc}
W(o) & \boldsymbol{0}\\
I & -\left(\Phi\left(o^{*}\right)\bar{L}_{u}\Phi\left(o^{*}\right)+I\right)
\end{array}\right].\label{eq:Jac}
\end{equation}

Note that for each equilibrium with its opinion formation $o^{*}$, when all of the opinion states \bs{are} evolving \bs{closely} enough to $o^{*}$, the gauge transformation matrix $\Phi\left(o^{*}\right)$ is fixed. 
\bs{From Remark 1, 
the spectrum of 
$-\left(\Phi\left(o^{*}\right) \bar{L}_{u}\Phi\left(o^{*}\right)+I\right)$ is 
the same as $- \left(\bar{L}_{u}+I\right)$.} Further, for 
\bs{any} opinions $o^{*}$, the matrices are Hurwitz. Hence, the stability of the system depends
on the \bs{spectrum} of $W\left(o\left(t\right)\right)$. \bs{From} \cite[Prop. 1]{bivirus}, \begin{align*}
s(-\left(D_{\min}+\left(D-D_{\min}\right)\left(O^{*}+0.5I\right)\right)\\
+\left(B-\left(O^{*}+0.5I\right)\left(B-B_{\min}\right)\right)) & <0
\end{align*}
if and only if \begin{align*}
\ensuremath{\rho(\left(D_{\min}+\left(D-D_{\min}\right)\left(O^{*}+0.5I\right)\right)^{-1}}\\
\times\left(B-\left(O^{*}+0.5I\right)\left(B-B_{\min}\right)\right))<1,
\end{align*}
Further, 
\bs{by} Proposition \ref{prop:Spe_r}, 
\begin{align*}
\ensuremath{\rho(\left(D_{\min}+\left(D-D_{\min}\right)\left(O^{*}+0.5I\right)\right)^{-1}}\\
\times\left(B-\left(O^{*}+0.5I\right)\left(B-B_{\min}\right)\right))\bs{\leq}R_{\max}\bs{\leq}1,
\end{align*}
we have $s\left(W\left(o\left(t\right)\right)\right)\bs{\leq}0$. \bs{Hence, the Jacobian matrices evaluated at healthy equilibria 
are Hurwitz if $R_{\max}<1$, leading to the results, by Lyapunov's indirect method.}
\qed
\vspace{2ex}

\textit{Proof of Theorem \ref{thm: stability_case1}:}
Note that 
\begin{align}
\dot{x}\left(t\right) &   =-\left[D_{\min}+\left(D-D_{\min}\right)\left(O\left(t\right)+0.5I\right)\right]x\left(t\right)\nonumber \\
 & \text{}+\left(I-X\left(t\right)\right)\left[B-\left(O\left(t\right)+0.5I\right)\left(B-B_{\min}\right)\right]x\left(t\right)\nonumber \\
 & = \text{}\text{\ensuremath{-D_{\min}x\left(t\right)+\left(I-X\left(t\right)\right)Bx\left(t\right)}}\nonumber \\
 & \text{}-\text{\ensuremath{\left(D-D_{\min}\right)\left(O\left(t\right)+0.5I\right)}\ensuremath{x\left(t\right)}}\nonumber\\
 &-\left(I-X\left(t\right)\right)\left(O\left(t\right)+0.5I\right)\left(B-B_{\min}\right)x\left(t\right)\nonumber \\
 & \leq\text{\ensuremath{-D_{\min}x\left(t\right)+\left(I-X\left(t\right)\right)Bx\left(t\right).}}\label{eq:sta}
\end{align}
The inequality implies further that
\[
\dot{x}\leq\dot{y}=-D_{\min}y(t)+By(t),
\]
since $I-X(t)$ is a diagonal matrix and $\left[I-X(t)\right]_{ii}\in\left[0,1\right]$, $\forall i\in\left[n\right]$.
From \cite[Prop. 1]{bivirus}, $R_{\max}=\rho(D_{\min}^{-1}B)\leq1$ implies
$s(-D_{\min}+B)\leq0$. 
\bs{Initializing} $y(0)=x(0)$, and from the fact that $\dot{y}=-D_{\min}y(t)+By(t)$
converges to $y=\boldsymbol{0}$ exponentially fast when $s(-D_{\min}+B)<0$, and $x=\boldsymbol{0}$ is the unique equilibrium when $R_{\max}<1$, we conclude that 
\bs{$x\left(t\right)\rightarrow\boldsymbol{0}$} exponentially
fast as {$t\rightarrow\infty$}. Additionally, 
since $(-D_{\min}+B)$ is an irreducible
Metzler matrix, if $s(-D_{\min}+B)=0$, \bs{by 
\cite [Lemma A.1]{khanafer2016SIS_positivesystems}, there exists a
positive diagonal matrix $P$ such that $(-D_{\min}+B)^{\top}P+P(-D_{\min}+B)$
is negative semidefinite.} Therefore, consider the Lyapunov function $V(x(t))=\frac{1}{2}x(t)^{\top}Px(t)$. From \eqref{eq:epi_c} and \eqref{eq:sta}, when $x(t)\ne\boldsymbol{0}$,
we have 
\begin{align*}
\dot{V}(x(t)) & =x(t)^{\top}P\dot{x}(t)\\
 & \le x(t)^{\top}P(-D_{\min}+B-X(t)B)x(t).
\end{align*}
Using the proof of 
\cite[Prop. 2]{bivirus},
one can show that $x(t)=\boldsymbol{0}$ is asymptotically stable
with the domain of attraction $[0,1]^{n}$.

\bs{Since $-(\bar{L}_{u}+I)$ is Hurwitz, based on Remark 1 and Theorem \ref{thm:equi_basic}, the convergence of each subsystem $\dot{o}(t)=-(\Phi\left(o^{*}\right)\bar{L}_u\Phi\left(o^{*}\right)+I)o(t)-0.5\boldsymbol{e}$ to $o^{*}$ is exponentially
fast under the fixed $\left(\Phi\left(o^{*}\right)\right)$. 
Further, each subsystem of the opinion dynamics in \eqref{eq:op_c} is input-to-state stable \cite[Lemma 4.6]{khalil2002nonlinear}. Therefore, 
the switching system in \eqref{eq:op_c} 
with input $x(t)$ vanishing to $\boldsymbol{0}$ asymptotically (or exponentially) fast, 
is asymptotically (or exponentially) stable with domain
of attraction $[-0.5,0.5]^{n}$.}
\qed

\vspace{2ex}
\textit{Proof of Theorem \ref{thm:case2}:}
Given the result of Lemma$~\ref{lem:set}$, it follows that the positive
invariance of $\Xi_{\epsilon}$ is established if we can prove that,
for all $i\in[n]$, 
$\dot{x}_{i}>0$ whenever $x_{i}=\epsilon y_{i}$
and $x_{j}\in[\epsilon y_{j},1]$ for $j\neq i$.
Toward that end, observe from \eqref{eq:epi} that 
\begin{align}
\label{eq:case2_pro}
\dot{x}_{i} & =-(\delta_{\min}(0.5-o_{i})+\delta_{i}\left(o_{i}+0.5\right))\epsilon y_{i}+(1-\epsilon y_{i})\nonumber \\
 & \quad\times\sum_{j\in \mathcal{N}_{i}}(\beta_{ij}(0.5-o_{i})+\beta_{\min}\left(o_{i}+0.5\right))(x_{j}-\epsilon y_{j}+\epsilon y_{j}),
\end{align}

\noindent
and note that we have dropped the argument $t$ for brevity. Since
$x_{j}-\epsilon y_{j}\geq0$ and $1>\epsilon y_{i}>0$ by hypothesis,
\[
(1-\epsilon y_{i})\sum_{j\in \mathcal{N}_{i}}(\beta_{ij}(0.5-o_{i})+\beta_{\min}\left(o_{i}+0.5\right))(x_{j}-\epsilon y_{j})\geq0.
\]

\noindent
This implies that 
\eqref{eq:case2_pro} obeys the following inequality:
\begin{align}
\label{eq:xi_2}
\dot{x}_{i} & \geq-(\delta_{\min}(0.5-o_{i})+\delta_{i}\left(o_{i}+0.5\right))\epsilon y_{i}\nonumber \\
 & \qquad+\sum_{j\in \mathcal{N}_{i}}(\beta_{ij}(0.5-o_{i})+\beta_{\min}\left(o_{i}+0.5\right))\epsilon y_{j}\nonumber \\
 & \qquad\quad-\epsilon^{2}y_{i}\sum_{j\in \mathcal{N}_{i}}(\beta_{ij}(0.5-o_{i})+\beta_{\min}\left(o_{i}+0.5\right))y_{j}.
\end{align}

\noindent
Note that $\phi y=(-D+B_{\min})y$ implies that $\delta_{i}y_{i}+\sum_{j\in \mathcal{N}_{i}}\beta_{\min}y_{j}=\phi_{i}y_{i}$.
Based on Assumption \ref{A1}, $\delta_{i}\geq\delta_{\min}$ and $\beta_{ij}\geq\beta_{\min}$
for $j\in \mathcal{N}_{i}$. \bs{Therefore}, we obtain 
\begin{equation}
-\delta_{\min}y_{i}+\sum_{j\in \mathcal{N}_{i}}\beta_{ij}y_{j}\geq-\delta_{i}y_{i}+\sum_{j\in \mathcal{N}_{i}}\beta_{\min}y_{j}=\phi_{i}y_{i}.\label{eq:dotx_3}
\end{equation}
\noindent
Since $o_{i}\in[-0.5,0.5]$, it follows \bs{from} \eqref{eq:dotx_3} \bs{that}
\begin{align}
 & \left(z_{i}+0.5\right)(-\delta_{i}y_{i}+\sum_{j\in \mathcal{N}_{i}}\beta_{\min}y_{j})\nonumber \\
 & \quad+(0.5-z_{i})(-\delta_{\min}y_{i}+\sum_{j\in \mathcal{N}_{i}}\beta_{ij}y_{j})\geq\phi_{i}y_{i}>0.
\end{align}

\noindent
Using \eqref{eq:xi}, the right-hand side of \eqref{eq:xi_2}
can then be further bounded as 
\[
\dot{x}_{i}\geq\epsilon\phi y_{i}-\epsilon^{2}y_{i}\sum_{j\in \mathcal{N}_{i}}(\beta_{ij}(0.5-z_{i})+\beta_{\min}\left(z_{i}+0.5\right))y_{j}.
\]

\noindent
 Obviously, for some sufficiently small $\epsilon_{i}>0$, we then
have
\[
\dot{x}_{i}\geq\epsilon_{i}\phi y_{i}-\epsilon_{i}^{2}y_{i}\sum_{j\in \mathcal{N}_{i}}(\beta_{ij}(0.5-z_{i})+\beta_{\min}\left(z_{i}+0.5\right))y_{j}>0.
\]

\noindent
By \bs{setting} \baike{$\bar{\epsilon}=\min_{i}\epsilon_{i}$}, we conclude that
$\Xi_{\epsilon}$ for every $\epsilon\in(0,\bar{\epsilon}]$ is a
positive invariant set of \eqref{eq:epi_op}.
Since $\Xi_{\epsilon}$, for $\epsilon\in(0,\bar{\epsilon}]$, is compact
and convex, 
the system in \eqref{eq:epi_op} is Lipschitz smooth in $\Xi_{\epsilon}$ under each switching subsystem in \eqref{eq:op_c}. 
Therefore, the result in \bs{\cite [Lemma 4.1]{yorke}} immediately establishes that any system
in \eqref{eq:epi_op} paired with one switching subsystem of \eqref{eq:op_c} has at least one equilibrium in $\Xi_{\epsilon}$. Taking
$\epsilon$ to be arbitrarily small, and 
Lemma \ref{lem:set} establishes
that the system in \eqref{eq:epi_op} has at least one equilibrium in \baike{ $\mathrm{Int}\chi$}. Therefore, the system in \eqref{eq:epi_op} can \bs{have} more than one endemic equilibrium.
\qed

\vspace{2ex}
\textit{Proof of Lemma \ref{lem:unstable}:}
Through the Jacobian matrix evaluated at $\left(\boldsymbol{0}, o^{*}\right)$ in \eqref{eq:Jac}, where the spectrum of $W\left(o\left(t\right)\right)$ determines the stability of the Jacobian matrix, 
\bs{it can be shown that} $ R_{\min}>1$ leads to,
\begin{align*}
s(-\left(D_{\min}+\left(D-D_{\min}\right)\left(O^{*}+0.5I\right)\right)\\
+\left(B-\left(O^{*}+0.5I\right)\left(B-B_{\min}\right)\right)) & >0,
\end{align*}
following 
the same 
process as in Proposition \ref{prop:unstable}. 
Therefore, the Jacobian matrix evaluated at $\left(\boldsymbol{0}, o^{*}\right)$ in \eqref{eq:Jac} has at least one positive eigenvalue. Hence, all the healthy equilibria $\left(\boldsymbol{0}, o^{*}\right)$, under the condition that $R_{\min}>1$, are unstable.
\qed
\vspace{2ex}

\textit{Proof of Theorem \ref{thm:case_3}:}
Under the condition that $R^{o*}_t<1$, proof of the local stability of all the healthy equilibria 
in Theorem \ref{thm:case_3} is similar to the proof of local stability of the healthy equilibria in Theorem \ref{thm: stability_case1}. By switching the condition $R_{\max}< 1$ to $R_{t}^{o^{*}}<1$, the Jacobian matrix in \eqref{eq:Jac} evaluated at $\left(\boldsymbol{0},o^{*}\right)$ is Hurwitz, 
which completes the proof of \bs{Case 1)} of the theorem. The proof of case 2) follows the same procedure under the condition that $R^{o*}_t>1$; \bs{showing} that $R_{t}^{o^{*}}>1$ 
\bs{implies} that the Jacobian matrix in \eqref{eq:Jac} evaluated at $\left(\boldsymbol{0},o^{*}\right)$ is not Hurwitz. \bs{Therefore,} 
the \bs{dissensus-healthy equilibria are unstable}. \bs{For Case 3), $R_{t}^{o^{*}}$ at 
$\left(\boldsymbol{0},-0.5\boldsymbol{e}\right)$ is 
$\rho\left(D\left(o_{\min}\right)^{-1}B\left(o_{\min}\right)\right)=R_{\max}$. Since $R_{\max}>1$ for moderate virus, from Lemma \ref{lem:unstable}}, the consensus-healthy equilibrium is unstable.
\qed

\vspace{2ex}
\textit{Proof of Corollary \ref{coro:method_2}:}
By selecting stubborn communities with opinion states fixed at $0.5$ from Corollary \eqref{coro:method_2}, based on Proposition \ref{prop:Spe_r}, the system in (\ref{eq:epi_op}) satisfies $R_{t}^{\vec{o}}<1$. 
Since all non-stubborn communities will have their opinion states greater or equal than $-0.5$, from Proposition \ref{prop:Spe_r}, $R^{o}_{t}<1$.
Thus, 
the system in \eqref{eq:epi_op} converges to a healthy state.\qed




\end{document}